\documentclass[lettersize,journal,twoside]{IEEEtran}
\usepackage{amsmath}

\usepackage{amssymb}
\usepackage{amsthm}
\usepackage{algorithm}
\usepackage[noend]{algpseudocode} 
\usepackage{array}
\usepackage{textcomp}
\usepackage{stfloats}
\usepackage{url}
\usepackage{verbatim}
\usepackage{graphicx}
\usepackage{cite}
\usepackage{paralist}
\usepackage{units}
\usepackage{booktabs} 
\usepackage{latexsym}
\usepackage{multirow}
\usepackage{xspace}
\usepackage{tikz}
\usepackage{caption}
\usepackage{subcaption}
\usepackage{float}
\usepackage{wrapfig}
\usepackage{stmaryrd}
\usepackage{wrapstuff}
\usepackage{threeparttable}

\usepackage[breaklinks]{hyperref}
\hypersetup{
colorlinks = true, 
linkcolor=blue, 
citecolor=blue, 
urlcolor=blue 
}

\usepackage{color}
\usepackage{colortbl}
\usepackage{tcolorbox}
\usepackage{mathrsfs}

\usepackage{pgf}
\usepackage{tikz}
\usetikzlibrary{arrows.meta,matrix,shapes,decorations,automata,backgrounds,positioning}

\usepackage{orcidlink}

\newcommand{\true}{\ensuremath{\mathtt{true}}\xspace}
\newcommand{\false}{\ensuremath{\mathtt{false}}\xspace}
\newcommand{\R}{\mathbb{R}}
\newcommand{\N}{\mathbb{N}}
\newcommand{\Rpos}{\R_{+}}
\newcommand{\Npos}{\N_{+}}

\newcommand{\RN}{\R^{N}}

\newcommand{\bw}{\ensuremath{\mathbf{w}}\xspace}

\newcommand{\Robust}[2]{\ensuremath{\mathsf{rob}(#1, #2)}\xspace}
\newcommand{\traceSat}[1]{\ensuremath{[\![#1]\!]^{\oplus}}\xspace}
\newcommand{\traceVio}[1]{\ensuremath{[\![#1]\!]^{\ominus}}\xspace}

\newcommand{\smallerSatEq}{\ensuremath{\preceq^{\oplus}}\xspace}
\newcommand{\smallerVioEq}{\ensuremath{\preceq^{\ominus}}\xspace}
\newcommand{\orderSat}{\ensuremath{\preceq^{\oplus}}\xspace}
\newcommand{\orderVio}{\ensuremath{\preceq^{\ominus}}\xspace}
\newcommand{\Var}{\mathbf{Var}}
\newcommand{\pair}[2]{\ensuremath{\left\langle#1, #2\right\rangle}\xspace}
\newcommand{\setVert}{\;\;\middle\vert\;\;}

\newcommand{\UntilOp}[1]{\mathbin{\mathcal{U}_{#1}}}

\newcommand{\Rnn}{\R_{\ge 0}}

\newcommand{\Defeq}{:=}

\newcommand{\STL}{\textrm{STL}}

\newcommand{\traceSet}{\ensuremath{W}\xspace}

\newcommand{\budget}{\ensuremath{\mathit{B}}\xspace}
\newcommand{\kset}[1]{\ensuremath{[#1]}\xspace}

\newcommand{\SCondPlus}{\ensuremath{\mathsf{Cls}^\oplus}\xspace}
\newcommand{\SCondMinus}{\ensuremath{\mathsf{Cls}^\ominus}\xspace}

\newcommand{\speed}{\ensuremath{\mathtt{speed}}\xspace}
\newcommand{\gear}{\ensuremath{\mathtt{gear}}\xspace}
\newcommand{\rpm}{\ensuremath{\mathtt{RPM}}\xspace}
\newcommand{\reaction}{\ensuremath{\mathtt{react}}\xspace}
\newcommand{\fals}{\ensuremath{\mathit{fals}}\xspace}

\newcommand{\targetone}{\ensuremath{\mathtt{task1}}\xspace}
\newcommand{\targettwo}{\ensuremath{\mathtt{task2}}\xspace}
\newcommand{\stable}{\ensuremath{\mathtt{stable}}\xspace}
\newcommand{\areaone}{\ensuremath{\mathtt{goal1}}\xspace}
\newcommand{\areatwo}{\ensuremath{\mathtt{goal2}}\xspace}
\newcommand{\danger}{\ensuremath{\mathtt{danger}}\xspace}
\newcommand{\acceleration}
{\ensuremath{\mathit{acceleration}}\xspace}

\newcommand{\brake}{\ensuremath{\mathtt{brake}}\xspace}

\newcommand{\af}{\ensuremath{\mathtt{AF}}\xspace}
\newcommand{\afref}{\ensuremath{\mathtt{AFref}}\xspace}

\newcommand{\valueSpace}{\ensuremath{\Theta}\xspace}
\newcommand{\nodes}{\ensuremath{\mathcal{N}}\xspace}
\newcommand{\edges}{\ensuremath{\mathcal{E}}\xspace}
\newcommand{\node}{\ensuremath{\nu}\xspace}
\newcommand{\edge}{\ensuremath{\epsilon}\xspace}
\newcommand{\graph}{\ensuremath{\mathcal{G}}\xspace}
\newcommand{\numClass}{\ensuremath{N}\xspace}
\newcommand{\numPath}{\ensuremath{P}\xspace}

\newcommand{\alwmid}{\texttt{AlwMid}\xspace}
\newcommand{\longbs}{\texttt{LongBS}\xspace}

\newcommand{\spec}[2]{\ensuremath{\varphi^{\mathsf{#1}}_{#2}}\xspace}

\newcommand{\myparagraph}[1]{\medskip\noindent{\bf #1.}}
\allowdisplaybreaks

\newcounter{myexample}
\newenvironment{myexample}[1][]{\refstepcounter{myexample}\par\medskip
 \textit{Example~\themyexample #1} \rmfamily}{\medskip}

 \newcounter{mylemma}
\newenvironment{mylemma}[1][]%
{%
\refstepcounter{mylemma}\par\smallskip
\noindent\textbf{Lemma~\themylemma}%
\if\relax\detokenize{#1}\relax\else\ (#1)\fi
\rmfamily
}
{\smallskip}

 \newcounter{mydefinition}
\newenvironment{mydefinition}[1][]%
{%
\refstepcounter{mydefinition}\par\smallskip
\noindent\textbf{Definition~\themydefinition}%
\if\relax\detokenize{#1}\relax\else\ (#1)\fi
\rmfamily
}
{\smallskip}

\newcounter{mycorollary}
\newenvironment{mycorollary}[1][]{\refstepcounter{mycorollary}\par\smallskip
\noindent \textbf{Corollary~\themycorollary #1} \rmfamily}{\smallskip}

\newcounter{myremark}

\newtheorem{mytheorem}{Theorem}
{\bfseries}{\itshape} 
{\bfseries}{\rmfamily}
{\bfseries}{\itshape}
{\bfseries}{\itshape}
{\bfseries}{\rmfamily}
\theoremstyle{definition}
{\bfseries}{\rmfamily}
{\bfseries}{\rmfamily}
{\bfseries}{\rmfamily}
{\bfseries}{\rmfamily}

\begin{document}

\title{Counterexample Classification for Signal Temporal Logic Specifications}

\author{

Zhenya Zhang~\orcidlink{0000-0002-3854-9846},
Parv Kapoor~\orcidlink{0000-0002-6093-0192},
Jie An~\orcidlink{0000-0001-9260-9697},
and Eunsuk Kang~\orcidlink{0000-0001-7891-6885}
\thanks{Zhenya Zhang is with Kyushu University, Fukuoka, Japan. Zhenya Zhang is also with the National Institute of Informatics, Tokyo, Japan. Email: zhang@ait.kyushu-u.ac.jp}
\thanks{Parv Kapoor and Eunsuk Kang are with Carnegie Mellon University, Pittsburgh, PA, USA. Email: \{parvk,eunsukk\}@andrew.cmu.edu}
\thanks{Jie An is with the Institute of Software, Chinese Academy of Sciences, Beijing, China. Email: anjie@iscas.ac.cn. }
}

\markboth{IEEE Transactions on Computer-Aided Design of Integrated Circuits and Systems,~Vol.~45,~No.~11, November~2026}%
{Z. Zhang~\MakeLowercase{\textit{et al.}}: Counteraxample Classification for Signal Temporal Logic Specifications}


\maketitle

\begin{abstract}
Signal Temporal Logic (STL) has been widely adopted as a specification language for specifying desirable behaviors of hybrid systems. One of the most common uses of STL is \emph{falsification}, which attempts to generate \emph{counterexample signals} that demonstrate how the system violates a given STL specification. A number of falsification methods and tools are available for efficient generation of counterexamples, which can be examined by the engineer to identify potential defects in the system. However, some of these counterexamples may be considered \emph{similar} to each other in that they describe system behavior that stems from the same underlying causes or defects. Since examining counterexamples can be a labor-intensive task, a tool that presents a distinct set of counterexamples and avoids showing repetitive ones could reduce the amount of effort that the engineer spends in debugging.

In this paper, we propose a  \emph{counterexample classification} method for STL specifications. Our approach is based on a novel criterion for classifying given counterexamples into a finite set of \emph{classes}, each of which corresponds to a set of signals that share a common behavioral pattern. In particular, each class is represented by a formula in \emph{parametric signal temporal logic (PSTL)}, which provides a concise description of the signals in the class; then, the problem of checking whether a given signal belongs to a particular class can be formulated as finding parameter values for the corresponding PSTL formula such that the signal satisfies the formula. Based on this idea, we propose an algorithm for automatically identifying classes from a given set of counterexamples. Since the algorithm needs to explore a potentially large space of classes, we also propose an efficient pruning method that leverages the concept of an \emph{inclusion relation} between different classes.
We have implemented a prototype tool of the proposed counterexample classification method. We demonstrate the efficiency of our algorithm and its utility on three hybrid systems, including automatic transmission, abstract fuel control, and robot navigation. 
\end{abstract}


\section{Introduction}\label{sec:intro}
Hybrid systems consist of both discrete events and continuous dynamics, and have been used as a mathematical model of cyber-physical systems (CPS). 
Quality assurance is a major concern in these systems, as many of them are deployed in safety-critical domains, such as automotive systems and medical devices, and misbehaviors of these systems can lead to catastrophic consequences. Moreover, the recent trend of integrating artificial intelligence (AI)-based components, such as vision-based controllers, has introduced new  safety risks. 

Different approaches have been developed to ensure the safety and reliability of hybrid systems. Among these approaches, \emph{formal verification}~\cite{chen13Flow,frehse2011spaceex} can bring rigorous guarantee on system safety via formal modeling and reasoning. However, as it essentially aims to cover whole state space, scaling verification to complex systems remains a challenge. \emph{Monitoring}~\cite{deshmukh2017robust,zhang2023online}, as a lightweight alternative, aims to assess specific system behaviors and detect safety violations at runtime, rather than explore the whole system state space. Based on runtime information, supplementary actions can be taken to ensure safety, such as runtime enforcement~\cite{su2025runtime}.  Another lightweight approaches are \emph{falsification}~\cite{donze2010breach,AnnpureddyLFS11,ARCHCOMP25Falsification}, which aims to find unsafe behaviors to expose system defects. Falsification has been vibrantly explored in the past decade, and the state-of-the-art approaches often adopt stochastic optimization to explore system state space efficiently.

\emph{Counterexamples}---executions that demonstrate how a system may violate a given  specification---play an important role in quality assurance.
In the context of hybrid systems, \emph{signal temporal logic (STL)}~\cite{maler2004monitoring} is often adopted as the specification language, and so counterexamples here refer to system trajectories (or \emph{signals}) that violate the given specification. As these trajectories can expose certain system defects, they are further inspected by the engineer to identify the root causes~\cite{bartocci2018localizing} and taken as references for system repair~\cite{lyu2025spectacle}.

\myparagraph{Motivation} While counterexamples are indispensable, inspecting and diagnosing them can be a time-consuming process for the system engineer. A typical system may yield a large (and possibly infinite) number of counterexamples, some of which are \emph{similar} to each other in that they belong to the same \emph{class} of counterexamples; i.e., they describe system behavior that stems from the same underlying causes or defects. Ideally, the engineer should be presented a diverse set of counterexamples from different classes, rather than having to examine  multiple signals from the same class.

\begin{figure}[!tb]
\centering
\includegraphics[width=0.8\columnwidth]{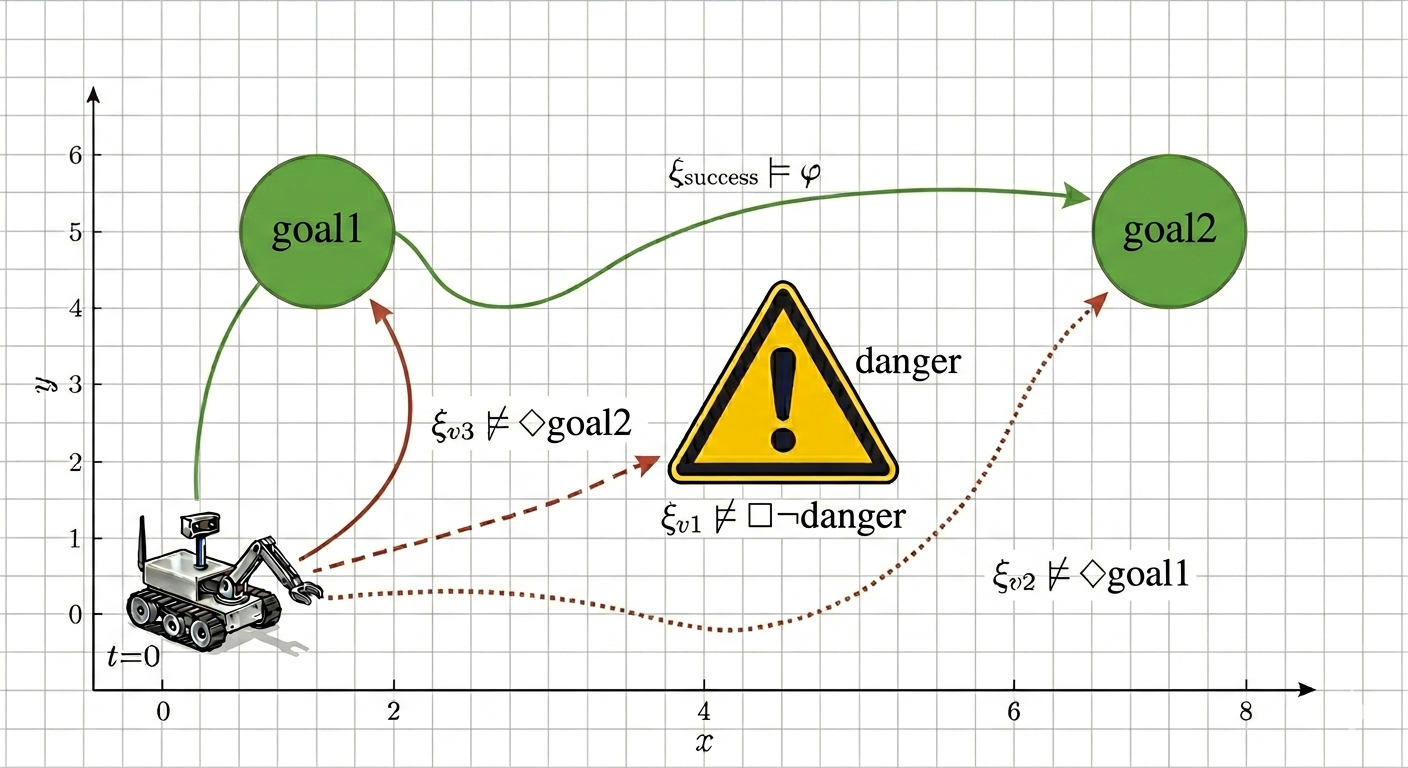}
\caption{A ``reach-avoid'' task for a robot}
\label{fig:robot}
\end{figure}
\begin{myexample}\label{ex:robotIntro}
    Fig.~\ref{fig:robot} shows an example in which a navigation robot performs a ``reach-avoid'' task, i.e, it should first reach goal area \areaone within time interval $I_1$; after that, it should reach \areatwo within $I_2$; during the whole interval $I$, it should avoid entering the dangerous area $\danger$. The requirement can be expressed by an STL formula:  $\varphi\equiv \Diamond_{I_1}(\areaone \land \Diamond_{I_2}(\areatwo)) \land \Box_{I}(\neg\danger)$, where $\Diamond$ and $\Box$ are respectively the \emph{eventually} and \emph{always} operators. 

    In this example, the robot may violate the specification in different ways. For example, it may fail to reach \areaone within $I_1$ or it may fail to reach \areatwo after reaching \areaone; it is also possible that it accomplishes both goals, but does not manage to avoid the dangerous area during the execution. These different violations are caused by different underlying causes. As the engineer attempts to debug and repair the system, being able to generate and explore these diverse violation scenarios is beneficial, as (1) it can reduce unnecessary effort in examining the same type of violation repeatedly and (2) it allows the engineer to prioritize which of the violations should be debugged and repaired first. \hfill$\lhd$
\end{myexample}

A number of tools and methods for efficient generation of counterexamples against STL specifications are available~\cite{ARCHCOMP25Falsification}. However, as far as we are aware, none of the existing methods specifically tackles the problem of \emph{classifying} counterexamples based on the behavior that they capture and \emph{automating} the process of generating different classes.

\begin{figure}[!tb]
\centering
\includegraphics[width=\columnwidth]{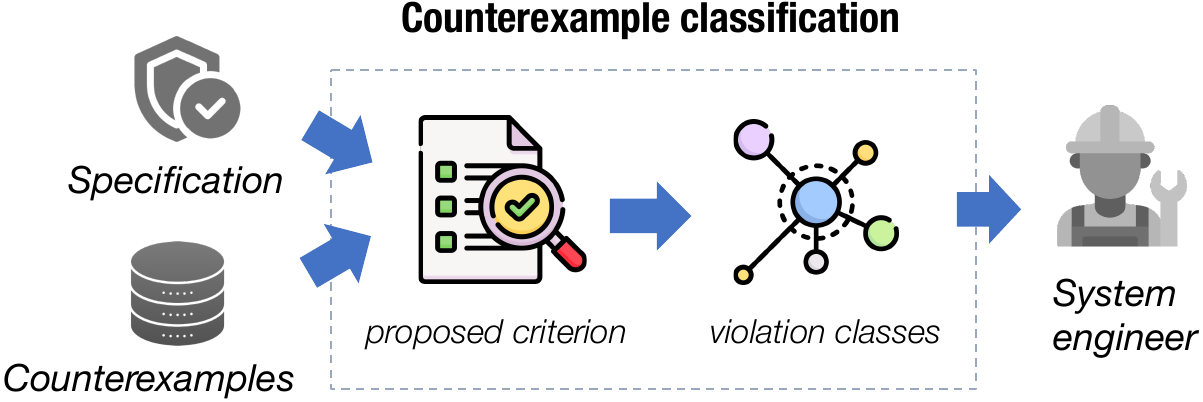}
\caption{Workflow of counterexample classification}
\label{fig:workflow}
\end{figure}

\myparagraph{Contribution} 
In this paper, we propose a novel \emph{counterexample classification framework for STL specifications}. 
Our framework is based on (i) a formal criterion for classifying  counterexamples against STL specifications and (ii) algorithmic methods for automatic, efficient generation of classes. Our approach is not tied to a specific counterexample generation method or tool; counterexamples may be collected by different approaches (e.g., random sampling or falsification), and as shown in Fig.~\ref{fig:workflow}, our framework can classify them into different classes, by which it exhibits the distribution of their shared behavioral patterns and helps shed  light on potential system defects.

We first propose a classification criterion in which each class signifies a \emph{sufficient condition} for a signal to satisfy or violate a given STL specification. The key intuition here is that there could be many different sufficient conditions to violate a specification, and each condition signifies a class that violates the specification in a specific way. 
We represent each class with a formula in \emph{parametric signal temporal logic (PSTL)}~\cite{asarin2012parametric}, because this formalism provides the expressivity to represent a group of signals and capture their patterns. 
We further formalize and prove the soundness of our criterion, i.e., each class indeed signifies such a sufficient condition. Note that by our criterion, each signal is not exclusively included in only one class; instead, it can be included in different classes due to the different features it may hold. Intuitively, the classes partition the traces of an STL formula to different sets, while our classification can capture the semantic meaning of these sets, each corresponding to a specific sufficient condition of satisfying or violating an STL formula.

Based on the criterion, we propose an automated counterexample classification method through a systematic traversal of  classes. Due to our selected formalism (i.e., PSTL), identifying the membership of a signal to a class requires searching for a valuation under which the signal violates the STL formula derived from the PSTL. This is known as a \emph{specification mining} problem and has been extensively studied in the literature~\cite{bartocci2022survey}. We adopt an existing approach proposed in~\cite{asarin2012parametric} for this purpose. 

However, a naive approach based on exhaustive class traversal is inefficient and unlikely to scale, given a possibly large number of classes. To mitigate this issue, we derive an \emph{inclusion relation} (i.e., a partial order) between different classes, which enables the classes to be described as a directed acyclic graph. For a given signal, in each path of this graph, there exists a boundary that separates the classes that include the signal and those that do not include the signal. Based on this idea, we formulate the problem of classifying a signal as that of finding the boundary in each path of the graph, which can be solved efficiently using a binary search algorithm. 

We have implemented our approach in a prototype tool and performed an experimental validation to evaluate the performance and effectiveness of our approach, based on two widely-adopted models (\emph{automatic transmission} and {abstract fuel control}) in the hybrid system community~\cite{ARCHCOMP25Falsification} as well as the robot navigation example. We show that our approach can accomplish classification tasks for a set of counterexamples against commonly-used specifications efficiently, thanks to the proposed inclusion relation. We also showcase some specific counterexamples and thereby demonstrate how our proposed criterion can differentiate different counterexamples in terms of their violation patterns.
These results can provide valuable insights for system debugging and reengineering, leading to a new step in the safety assurance of hybrid systems.

\section{Preliminaries: Signal Temporal Logic}\label{sec:preliminary}

Let $T\in \Rpos$ be a positive real. An \emph{$N$-dimensional time-variant signal} with a time horizon $T$ is defined as a function $\bw\colon [0,T]\to\R^{N}$. Signals are used to represent the execution traces of hybrid systems, and each dimension of a signal is associated with a \emph{signal variable} which has a specific physical meaning, such as \speed, \rpm and \acceleration of an automotive system. 

Signal temporal logic (STL)~\cite{maler2004monitoring} extends \emph{linear temporal logic (LTL)} to continuous temporal and spatial domains, and has been widely adopted to specify behaviors of hybrid systems. 
Below, we outline the syntax and semantics of STL.

\begin{mydefinition}[STL Syntax]\label{def:stlSyntax}
We fix a set $\Var$ of variables. In STL, \emph{atomic propositions} and  \emph{formulas} are defined as follows:
\begin{math}
      \alpha 
 \,::\equiv\,
        f(x_1, \dots, x_N) > 0, \;
      \varphi  \,::\equiv\,
        \alpha \mid \bot
        \mid \neg \varphi 
        \mid  \varphi_1\land\varphi_2
        \mid  \varphi_1\lor \varphi_2
        \mid \Box_{I}\varphi
        \mid  \Diamond_{I}\varphi
        \mid \varphi \UntilOp{I} \varphi
\end{math}.
Here
 $f$ is an $N$-ary function $f:\RN \to \R$, $x_1, \dots, x_N \in \Var$,
  and $I$ is a closed non-singular interval in $\Rnn$,
  i.e.,\ $I=[a,b]$, where $a,b \in \Rpos$ and $a<b$.
$\Box, \Diamond$ and $\mathcal{U}$ are temporal operators, which are known as \emph{always}, \emph{eventually} and \emph{until} respectively. 
The always operator $\Box$ and eventually operator $\Diamond$ can be considered as special cases of the until operator $\mathcal{U}$, and they can be derived by until operator as follows: $\Diamond_{I}\varphi\equiv\top\UntilOp{I}\varphi$ and $\Box_{I}\varphi\equiv\lnot\Diamond_{I}\lnot\varphi$.
Other common predicates such as $\rightarrow,\top$  are introduced as syntactic sugar: $\top \equiv\neg\bot$, $\varphi_1\to\varphi_2 \equiv \neg\varphi_1 \lor \varphi_2$. 
\end{mydefinition}

\begin{mydefinition}[Robust Semantics]\label{def:robSemantics} 
Let $\bw \colon [0,T]\to \R^{N}$ be an $N$-dimensional signal, and let $\varphi$ be an $\STL$ formula. The robust semantics of $\varphi$ returns $\Robust{\bw}{\varphi} \in \R \cup \{\infty,-\infty\}$, called \emph{robustness}, as follows, by induction on the construction of formulas.
\begin{align*}
 &\Robust{\bw}{\alpha} \;\Defeq \;
 f\bigl(\bw(0)(x_1), \cdots, \bw(0)(x_N)\bigr) 
 \hspace{1.5em}\\
 &\Robust{\bw}{\bot} \;\Defeq\; -\infty
 \\ &\Robust{\bw}{\neg \varphi} \;\Defeq\; - \Robust{\bw}{\varphi}
 \\
 &\Robust{\bw}{\varphi_1\land \varphi_2} \;\Defeq\; \min\left({\Robust{\bw}{\varphi_1}, \Robust{\bw}{\varphi_2}}\right) \\
 &\Robust{\bw}{\varphi_1\lor\varphi_2} \;\Defeq\; \max\left(\Robust{\bw}{\varphi_1}, \Robust{\bw}{\varphi_2}\right)\\
 &\Robust{\bw}{\Box_{I}\varphi} \;\Defeq \; \inf_{t\in I}{\left(\Robust{\bw^t}{\varphi}\right)} \\
  &\Robust{\bw}{\Diamond_{I}\varphi} \;\Defeq \; \sup_{t\in I}{\left(\Robust{\bw^t}{\varphi}\right)}
 \\
&\Robust{\bw}{\varphi_1 \UntilOp{I} \varphi_2} \Defeq
 { \sup_{t \in I}\min \left( 
 \begin{array}{l}
   \Robust{\bw^t}{\varphi_2},    \\
    \inf_{t' \in [0, t)} \Robust{\bw^{t'}}{\varphi_1}
 \end{array}
 \right)}
\end{align*}
where $\bw^t\colon [0,T-t]\to \R^{N}$ denotes the $t$-shift of $\bw$, i.e., the signal $\bw^t$ is defined by $\bw^t(t') \Defeq \bw(t+t')$. 
\end{mydefinition}

Essentially, the robust semantics in Def.~\ref{def:robSemantics} refines the Boolean semantics, and we can derive it as follows: $ \Robust{\bw}{\varphi} > 0$ implies $\bw\models\varphi$, and $\Robust{\bw}{\varphi} < 0$ implies $\bw\not\models\varphi$. See~\cite[Prop.~16]{fainekos2009robustness} for more details.
In the following, we denote by $\traceSat{\varphi}$ the set of  signals that satisfy $\varphi$, i.e., $\traceSat{\varphi} \Defeq \{\bw\mid \bw\models\varphi\}$; and denote by $\traceVio{\varphi}$ the set of signals that violate $\varphi$, i.e., $\traceVio{\varphi} \Defeq \{\bw\mid \bw\not\models\varphi\}$.

\medskip
Given an STL formula $\varphi$, a \emph{counterexample} of $\varphi$ is a signal $\bw$ violating $\varphi$, i.e., $\bw \in \traceVio{\varphi}$. A signal $\bw$ can be a system output, so the existence of a counterexample $\bw \in \traceVio{\varphi}$ implies that the system can violate the specification $\varphi$. In practice, for a given specification $\varphi$, there could be many counterexamples arising from different causes and, thus, reflecting different system defects. In light of this, we aim to classify these counterexamples into different classes.

\section{Classification Criterion and A Straightforward Classification Approach}\label{sec:criteria}
We first introduce our proposed classification criterion in~\S{}\ref{sec:criterionSubsec}, and then a straightforward classification approach directly derived from the criterion in~\S{}\ref{sec:naiveApproach}.

\subsection{Classification Criterion}\label{sec:criterionSubsec}

\myparagraph{Parametric Signal Temporal Logic}\label{sec:paramSTL}
PSTL was introduced in~\cite{asarin2012parametric} and can be used as templates for mining STL specifications. The idea of PSTL is to parameterize the constants in normal STL, such that a PSTL formula induces a parameter valuation space, in which each valuation identifies a normal STL formula. 
While PSTL can have both \emph{magnitude parameters} (i.e., by parameterizing the constant $0$ in atomic propositions $\alpha$ in Def.~\ref{def:stlSyntax}) and \emph{timing parameters} (i.e., by parameterizing the intervals in temporal operators), in Def.~\ref{def:pstl} we adopt the fragment of PSTL that concerns timing parameters only, to showcase the underlying idea of classifying counterexamples; meanwhile, our criterion is extensible by taking magnitude parameters into account, leading to a multi-granularity classification approach customizable based on users' needs.

\begin{mydefinition}[Timing PSTL]\label{def:pstl}
 In timing PSTL, an \emph{atomic proposition} $\alpha$ is defined in the same way as that in Def.~\ref{def:stlSyntax}, and a PSTL \emph{formula} $\psi$ is defined as follows:
\begin{math}
\psi  \,::\equiv\,
        \alpha \mid \bot
        \mid \neg \psi 
        \mid  \psi_1\land\psi_2
        \mid  \psi_1\lor \psi_2
        \mid \Box_{[l, u]}\psi
        \mid  \Diamond_{[l,u]}\psi
        \mid \psi \UntilOp{[l, u]} \psi
\end{math},
where $l$ and $u$ are variables (parameters) that range in the space of positive reals, and hold the constraint that $l<u$.
Given a PSTL formula $\psi$ with $m$ parameters, the parameters in $\psi$ induce a space $\valueSpace \subseteq \Rpos^m$ of their valuations. By fixing a valuation $\theta\in \valueSpace$, we can transform $\psi$ into a normal STL formula $\varphi$, which we denote as $\varphi = \psi(\theta)$. In this sense, an STL formula can be seen as a special PSTL formula, where the valuation space only contains a specific valuation. 
\end{mydefinition}

The semantics of a PSTL formula $\psi$ given in~\cite{asarin2012parametric} concerns with the valid space of the valuation $\theta$ for a given signal $\bw$, such that the STL formulas $\psi(\theta)$ derived by applying $\theta$ to $\psi$ can make $\bw$ satisfy $\psi(\theta)$. For details about the computation of such valid space, we refer readers to~\cite{asarin2012parametric}. 

\myparagraph{Satisfiable/Violable set} We consider an extended concept from the original semantics of PSTL, called \emph{satisfiable set} (and dually, \emph{violable set}), which characterizes the set of signals that are possible to satisfy or violate a PSTL formula $\psi$, in the sense that they can satisfy or violate an STL formula $\psi(\theta)$ derived by assigning a valuation $\theta$ to $\psi$. Intuitively, as later we will use a PSTL formula $\psi$ to identify a class, the signals included in class $\psi$ are the satisfiable/violable set of $\psi$.

\begin{mydefinition}[Satisfiable/Violable Sets]\label{def:volume} 
Given a PSTL formula $\psi$, a \emph{satisfiable set} $\traceSat{\psi}$ of $\psi$ is the set of signals $\bw$ for which there exists $\theta$ such that $\bw\models\psi(\theta)$, i.e., $\traceSat{\psi} \Defeq \{\bw\mid \exists\theta\in\valueSpace.\;\bw\models\psi(\theta)\}$. 
Similarly, a \emph{violable set} $\traceVio{\psi}$ of $\psi$ is the set of signals $\bw$ for which there exists $\theta$ such that $\bw\not\models\psi(\theta)$, i.e., $\traceVio{\psi} \Defeq \{\bw\mid \exists\theta\in\valueSpace.\;\bw\not\models\psi(\theta)\}$.
\end{mydefinition}

Note that, by definition, it is possible that $\bw \in \traceSat{\psi}$ and $\bw \in \traceVio{\psi}$ because of different valuations of $\theta$, which is different from the case where $\psi$ is an STL formula.

\begin{table*}[!tb]
    \caption{The formal definitions of satisfaction classes $\SCondPlus(\varphi)$ and violation classes $\SCondMinus(\varphi)$}
    \label{tab:criterion}
    \centering
        \begin{align*}
         \midrule
        &\SCondPlus(\alpha) \;\Defeq\; \{\bot, \alpha\} \qquad
        \SCondPlus(\bot) \;\Defeq\; \{\bot\} \qquad
        \SCondPlus(\neg\varphi) \;\Defeq\; \left\{\neg\psi \mid \psi \in \SCondMinus(\varphi)\right\} \\
        &\SCondPlus(\varphi_1\land \varphi_2) \;\Defeq\; \left\{ \psi_1\land\psi_2 \mid \psi_1 \in \SCondPlus(\varphi_1), \psi_2\in\SCondPlus(\varphi_2)\right\} \qquad
         \SCondPlus(\varphi_1\lor \varphi_2) \;\Defeq\; \left\{\psi_1\lor\psi_2\mid \psi_1\in\SCondPlus(\varphi_1), \psi_2\in \SCondPlus(\varphi_2) \right\}
               \\
        & \SCondPlus(\Box_{[a, b]}\varphi) \;\Defeq\; \left\{ \bigwedge_{i\in\kset{k}}\Box_{[l_i, u_i]}\psi_i \mid \psi_i \in \SCondPlus(\varphi)
        \right\}  \qquad \SCondPlus(\Diamond_{[a, b]}\varphi) \;\Defeq\; \left\{ \bigvee_{i\in \kset{k}}\Diamond_{[l_i, u_i]}\psi_i \mid \psi_i \in \SCondPlus(\varphi)
        \right\} \\
        &\SCondPlus(\varphi_1\UntilOp{[a,b]}\varphi_2) \;\Defeq\; 
\left\{
      \exists i \in \kset{k}
      \Bigl(
        \psi_i \UntilOp{[l_i,u_i]} \sigma_i
        \land
        \forall j<i,\,
        \Box_{[l_{j},u_{j}]} \xi_{j}
      \Bigr) \;\;\middle | \;\;
         \psi_i,\xi_{j} \in \SCondPlus(\varphi_1), 
       \sigma_i \in \SCondPlus(\varphi_2)
\right\}
           \\\midrule
        &\SCondMinus(\alpha) \;\Defeq\; \{\top, \alpha\} \qquad
        \SCondMinus(\top) \;\Defeq\; \{\top\} \qquad
        \SCondMinus(\neg\varphi) \;\Defeq\; \left\{\neg\psi \mid \psi \in \SCondPlus(\varphi)\right\} \\
         &\SCondMinus(\varphi_1\land \varphi_2) \;\Defeq\; \left\{ \psi_1\land\psi_2 \mid \psi_1\in\SCondMinus(\varphi_1), \psi_2\in \SCondMinus(\varphi_2) \right\}
              \qquad
         \SCondMinus(\varphi_1\lor \varphi_2) \;\Defeq\; 
            \left\{ \psi_1\lor \psi_2 \mid \psi_1 \in \SCondMinus(\varphi_1), \psi_2\in\SCondMinus(\varphi_2)\right\} \\
        &\SCondMinus(\Box_{[a,b]}\varphi) \;\Defeq\; \left\{ \bigwedge_{i\in \kset{k}} \Box_{[l_i, u_i]}\psi_i \mid \psi_i \in \SCondMinus(\varphi)
        \right\} 
        \qquad
        \SCondMinus(\Diamond_{[a,b]}\varphi) \;\Defeq\; \left\{ \bigvee_{i\in\kset{k}}\Diamond_{[l_i, u_i]}\psi_i \mid \psi_i \in \SCondMinus(\varphi)
        \right\} \\
        & \SCondMinus(\varphi_1\UntilOp{[a,b]}\varphi_2) \Defeq  \left\{
      \exists i \in \kset{k}
      \Bigl(
        \psi_i \UntilOp{[l_i,u_i]} \sigma_i
        \land
        \forall j<i,\,
        \Box_{[l_{j},u_{j}]} \xi_{j}
      \Bigr) \;\;\middle | \;\;
         \psi_i,\xi_j \in \SCondMinus(\varphi_1), 
       \sigma_i \in \SCondMinus(\varphi_2)
\right\} \\ \midrule
        \end{align*}
\end{table*}
\myparagraph{Classification criterion} 
Based on Def.~\ref{def:volume}, we define our classification criterion. Intuitively, to classify the counterexamples of a given STL specification, we need to split the set of counterexamples into a number of subsets, each of which can identify a \emph{sufficient condition} to violate the given specification, because being included in a subset implies being included in the whole counterexample set. However,  classifying counterexamples for a given STL systematically and automatically, remains  challenging, due to the complexity of STL syntax and counterexamples' temporal behaviors. 

In Def.~\ref{def:classCriterion}, we propose a criterion which can systematically classify the counterexamples of a given STL formula. Notably, our criterion requires a set of user-specified parameters $k\in \Npos$ for each temporal operator (i.e., $\Box$, $\Diamond$ and $\UntilOp{}$) in the formula, thereby we can split the intervals of the temporal operators and predicate over each segment. For simplicity, we use $\kset{k}$ to denote the set of integers $\{1,\ldots,k\}$.

\begin{mydefinition}[Classification Criterion] \label{def:classCriterion}
Let $\varphi$ be an STL formula.  
For each temporal operator with a closed time interval $[a, b]$, 
we assign a positive integer $k\in\Npos$ by which we  split  $[a, b]$ into $k$ segments $[l_i, u_i]$ ($i\in\kset{k}$), such that, $l_1 = a$, $u_k = b$, and $(u_{i} = l_{i+1})$ for any $i\in\kset{k-1}$. Note that, the $k-1$ breakpoints $u_{i}$ are  the parameters of the PSTL formulas, each identifying a class; each parameter $u_{i}$ ranges in $[a,b]$, and the valuation space for these parameters is defined as $\valueSpace \Defeq \{\langle u_1,\ldots, u_{k-1} \rangle\in [a,b]^{k-1}\mid \forall i\in\kset{k-1}.(u_i < u_{i+1})\}$. 

The formal definition of our classification criterion is shown in Table~\ref{tab:criterion}. 
$\SCondPlus(\varphi)$ returns a set of PSTL formulas, in which each PSTL formula $\psi\in \SCondPlus(\varphi)$ is called a \emph{satisfaction class}. A signal $\bw$ is included in a class $\psi\in \SCondPlus(\varphi)$ if it satisfies $\bw\in\traceSat{\psi}$, i.e., $\bw$ is included in the satisfiable set of $\psi$. 

Dually, $\SCondMinus(\varphi)$ returns a set of PSTL formulas, in which each PSTL formula $\psi\in\SCondMinus(\varphi)$ is called a \emph{violation class}. A signal $\bw$ is included in a class $\psi\in \SCondMinus(\varphi)$ if it satisfies $\bw\in\traceVio{\psi}$, i.e., $\bw$ is included in the violable set of $\psi$. \qed
\end{mydefinition}

 Note that, although we mainly aim at classification of counterexamples, since the definition of counterexample classes relies on the dual case, i.e., the satisfaction classes, Def.~\ref{def:classCriterion} includes both cases, thus not only applicable to counterexamples but also to signals that satisfy the given specification. Intuitively, depending on the semantics of different operators, Def.~\ref{def:classCriterion} classifies their counterexamples in different ways:
\begin{compactitem}
    \item {\it Atomic proposition $\alpha$}: Def.~\ref{def:classCriterion} adopts the simple strategy, namely, it does not split the satisfiable/violable set of $\alpha$. While splitting that for $\alpha$ is possible (e.g., by considering magnitude parameters of PSTL), as our main aim is to show how our criterion works for Boolean connectives and temporal operators, we adopt the simplest strategy. 
    \item {\it Negation}: Intuitively, the satisfaction classes of $\neg\varphi$ is relevant to the violation classes of $\varphi$, and vice versa, because the signals that satisfy $\neg\varphi$ must violate $\varphi$.
    \item {\it Boolean connectives}: The criteria for $\land$ and $\lor$ depend on their semantics. For example, given a conjunction $\alpha_1\land \alpha_2$ of atomic propositions, the counterexample of this conjunction can arise from the violation of $\alpha_1$ and can also arise from the violation of $\alpha_2$; therefore, both $\alpha_1$ and $\alpha_2$ are a class in $\SCondMinus(\alpha_1\land \alpha_2)$ (in this case, $\SCondMinus(\alpha_1\land \alpha_2) = \{\top, \alpha_1, \alpha_2, \alpha_1\land\alpha_2\}$). On the other hand, a disjunction $\varphi = \alpha_1\lor \alpha_2$ can only be violated by violating $\alpha_1$ and $\alpha_2$ together, so $\alpha_1\lor \alpha_2$ is a class in  $\SCondMinus(\alpha_1\lor \alpha_2)$, but a single $\alpha_1$ or $\alpha_2$ is not.
    \item {\it Temporal operators}: To handle a temporal operator, we split its time interval into a number of segments, such that we can treat the temporal operator as the conjunction or disjunction of those segments and thereby we can handle it in the same way as we do for Boolean connectives. For example, given a formula $\Box_{[a,b]}\alpha$, by splitting its time interval $[a, b]$ into $[a,c]$ and $[c,b]$, we can rewrite the formula as $\Box_{[a,c]}\alpha \land \Box_{[c,b]}\alpha$; then in line with conjunction, both $\Box_{[a,c]}\alpha$ and $\Box_{[c,b]}\alpha$ are the violation classes, suggesting that the formula can be violated in either way.
\end{compactitem}

We remark that, Def.~\ref{def:classCriterion} achieves the recursive definition of the criterion via the following featured technical designs:
\begin{compactitem}
    \item {\it Usage of $\bot$ and $\top$.} By definition, the satisfiable set $\traceSat{\bot}$ and the violable set $\traceVio{\top}$ are both empty; nevertheless, these two predicates are useful in unifying the formalization of the criteria for different operators. For example, given the formula $\alpha_1\land \alpha_2$, then $\alpha_1$ identifies a counterexample class; this can be expressed by Def.~\ref{def:classCriterion} in which $\SCondMinus(\alpha_1\land \alpha_2) = \{ \psi_1\land\psi_2$\}, with $\psi_1 = \alpha_1$ and $\psi_2 = \top$; on the other hand, $\traceVio{\alpha_1}$ is not a counterexample class of $\alpha_1\lor \alpha_2$, and consistently, $\SCondMinus(\alpha_1\lor\alpha_2)$ does not contain $\alpha_1$ as a counterexample class by Def.~\ref{def:classCriterion}.
    \item {\it Finite split for temporal operators.} Intuitively, temporal operators are either existential (i.e., $\Diamond$) or universal (i.e., $\Box$); by splitting their intervals into a finite number of segments, we can align their classification criteria with Boolean connectives. Consequently, this allows us to reason about temporal orders of different events; one such an example is given in Example~\ref{ex:robotTech}. Moreover, by setting different $k$, we can also adjust the granularity of our classification, as showcased in~\S{}\ref{sec:evaluation}.
\end{compactitem}

\myparagraph{Soundness and completeness} As mentioned, each PSTL formula $\psi$ in $\SCondPlus(\varphi)$ or $\SCondMinus(\varphi)$ signifies a \emph{sufficient condition} for the satisfaction or violation of $\varphi$, in the following sense: given $\psi\in \SCondMinus(\varphi)$, if a signal $\bw$ is included in the satisfiable set of $\psi$, i.e., it can satisfy $\psi(\theta)$ under some valuation $\theta$, then it implies that $\bw$ satisfies $\varphi$; given $\psi\in \SCondMinus(\varphi)$,  if a signal $\bw$ is included in the violable set of $\psi$, i.e., it can violate $\psi(\theta)$ under some valuation $\theta$, then it implies that $\bw$ violates $\varphi$. 
Specially, $\bot$ and $\top$ serve as empty sets $\traceSat{\bot}$ and $\traceVio{\top}$ while holding the property of being as sufficient conditions. 
The intuition is formally stated  Theorem~\ref{lem:sound} and proved in~Appendix~\ref{app:Theo1}.

\begin{mytheorem}
    [Soundness]\label{lem:sound}
    Given any $\psi\in \SCondPlus(\varphi)$, if $\bw\in \traceSat{\psi}$ (i.e., it holds that $\exists\theta\in \valueSpace.\;\bw \models \psi(\theta)$), then it holds that $\bw \models \varphi$; given any $\psi \in \SCondMinus(\varphi)$, if $\bw\in \traceVio{\psi}$ (i.e., it holds that $\exists\theta\in \valueSpace.\;\bw \not\models\psi(\theta)$), then it holds that $\bw \not\models\varphi$.
\end{mytheorem}

Intuitively, the completeness of a classification criterion refers to the property that any counterexample $\bw$ of a given STL formula $\varphi$ must be included in one of the classes of $\varphi$. In our case, by Def.~\ref{def:classCriterion}, as any $\varphi$ is included as a class of its own (i.e., $\varphi \in \SCondPlus(\varphi)$ and $\varphi \in \SCondMinus(\varphi)$), it ensures our criterion to hold completeness naturally. More importantly, this is also useful in practice, because it also ensures that our criterion is complete despite the setting of $k$. Given a temporal operator, while the setting of $k$ allows us to match different classes of its sub-formula $\varphi$ in different intervals, as it holds $\varphi \in \SCondMinus(\varphi)$, we can always match $\varphi$ as the class in any interval. In Theorem~\ref{lem:complete}, we state this completeness; see~Appendix~\ref{app:theo2} for its proof.

\begin{mytheorem}[Completeness]\label{lem:complete}

 Given an STL formula $\varphi$, despite the setting of $k$ in Def.~\ref{def:classCriterion}, for any $\bw$ such that $\bw \models \varphi$,  there exists  $\psi\in\SCondPlus(\varphi)$ such that $\bw\in\traceSat{\psi}$, and for any $\bw$ such that $\bw \not\models \varphi$, there exists $\psi\in\SCondMinus(\varphi)$ such that $\bw\in\traceVio{\psi}$. 
\end{mytheorem}

Moreover, we also give the following lemma which states that the classes of a formula form a closure under $\lor$ or $\land$ and sketch the proof in~Appendix~\ref{app:Lem1}. Later this lemma is useful in the proof of Theorem~\ref{theo:relationComplete}.

\begin{mylemma}\label{lem:closure}
  Given an STL formula $\varphi$, $\SCondPlus(\varphi)$ is closed under $\lor$, and $\SCondMinus(\varphi)$ is closed under $\land$.  
\end{mylemma}

\begin{myexample}\label{ex:robotTech}
 We revisit the robot example in Example~\ref{ex:robotIntro}, where $\varphi\equiv \Diamond_{I_1}(\areaone \land \Diamond_{I_2}(\areatwo)) \land \Box_{I}(\neg\danger)$. By applying our criterion in Def.~\ref{def:classCriterion} (with $k = 1$ for each temporal operator), we can obtain 8 violation classes of $\varphi$ (including the universal $\varphi$ and the empty $\top$); we look into the following three classes each of which is an element of $\SCondMinus(\varphi)$ and identifies a distinct cause that can explain the violation of $\varphi$ from different aspects:
 \begin{align*}
     \Diamond_{I_1}(\areaone)&\text{: failed to reach \areaone in time} \\
     \Diamond_{I_1}(\Diamond_{I_2}\areatwo)&\text{: failed to reach \areatwo in time}\\
     \Box_{I}(\neg\danger)&\text{: entered the dangerous zone at some point}
 \end{align*}
 Moreover, by setting $k = 2$ for $I_1$, we can obtain more classes that characterize the causal reasons more finely. We take the following two classes for an explanation (below we let $I_1 = [0, 100]$ and use $t$ as a parameter). Note that, both classes are counterexample classes (i.e., an element of $\SCondMinus(\varphi)$), so a counterexample in their violable set should violate their instantiated STL formulas and holds an intuitive explanation as stated following each formula below.
\begin{equation}
\begin{aligned}
     &\Diamond_{[0, t]}(\areaone) \lor \Diamond_{[t, 100]}(\Diamond_{I_2}(\areatwo)) \label{class:1}\\
      &\qquad \text{\areaone missed firstly, and \areatwo missed later}
\end{aligned}
\end{equation}
\begin{equation}
\begin{aligned}
     &\Diamond_{[0, t]}(\Diamond_{I_2}(\areatwo))\lor \Diamond_{[t, 100]}(\areaone) \label{class:2} \\
      &\qquad \text{\areatwo missed firstly, and \areaone missed later}
\end{aligned}
\end{equation}
 For example, Class~(\ref{class:1}) can describe a trajectory in which the robot makes the order of the two goals wrong, i.e., it first reaches \areatwo and then reaches \areaone. By contrast, Class~(\ref{class:2}) can describe another situation, where the robot first reaches \areaone successfully, but fails to reach \areatwo in time later.
 \hfill$\lhd$
\end{myexample}

\begin{figure}[!t]
    \centering
    \begin{subfigure}{0.45\columnwidth}
        \centering
        \includegraphics[width=\linewidth]{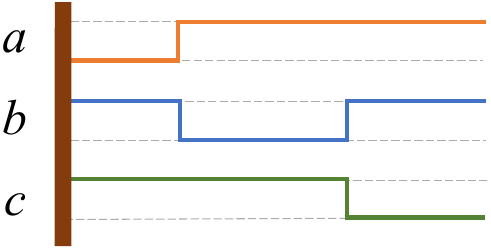}
        \caption{}\label{fig:highK1}
    \end{subfigure}
    \quad
    \begin{subfigure}{0.45\columnwidth}
        \centering
        \includegraphics[width=\linewidth]{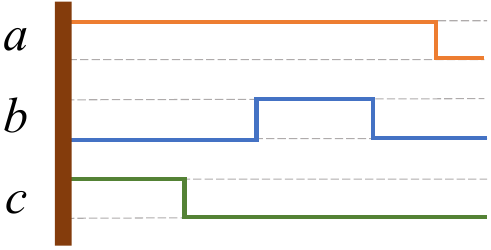}
        \caption{}\label{fig:highK2}
    \end{subfigure}
    \caption{Two example signals that exhibit different orders of events that lead to specification violation}
    \label{fig:highKExample}
\end{figure}
\begin{myexample}\label{ex:highK}
We consider an example of a UAV system. The specification $\Diamond_{[0,T]}(a \land b \land c)$ requires that three conditions $a$, $b$ and $c$ are satisfied simultaneously at some point, where $a$ denotes available GPS localization, $b$ denotes valid path planning, and $c$ denotes autopilot execution. Though not explicitly encoded in the specification, the intended operation of the UAV in practice, assumes that $a$, $b$, and $c$ occur in sequence: GPS localization is established first, path planning is then completed, and the autopilot is finally executed.

In this system, specification violation in terms of different orders of events may arise from different causes. Fig.~\ref{fig:highKExample} illustrates two possible execution trajectories, which suggest two different causes. Fig.~\ref{fig:highK1} illustrates an order of violation $a \to b \to c$; in this case, GPS first fails to localize, and because of this, the planning module cannot propose a meaningful path, and therefore the autopilot cannot be executed. In contrast, Fig.~\ref{fig:highK2} demonstrates another order $b\to c \to a$, which suggests a different cause. In this case, the GPS works normally, however, the planning module fails to make a path planning, and consequently, the autopilot does not work normally; although GPS also fails later, it is not the cause of the failure -- the real cause is the planning module.  

By setting $k = 3$, we can capture the different violation patterns. In this example, Fig.~\ref{fig:highK1} can be captured by $\Diamond_{[0,t_1]}a \lor \Diamond_{[t_1, t_2]}b \lor \Diamond_{[t_2, T]}c$, i.e., $a$ is violated firstly, followed by $b$ and $c$. Instead, Fig.~\ref{fig:highK2} can be captured by another class $\Diamond_{[0,t_1]}b \lor \Diamond_{[t_1, t_2]}c \lor \Diamond_{[t_2, T]}a$, where the violation of $a$ happens after the violations of $b$ and $c$. By distinguishing different patterns of a specification violation, our criterion provides useful diagnostic insights into the causes of system failures.
\end{myexample}

\myparagraph{Selection of $k$} 
Selecting a proper $k$ is important to capture the information hidden in counterexamples.
In Example~\ref{ex:highK}, while we have seen that setting $k = 3$ is useful to capture the cause of the violation, setting $k= 2$ captures only part of the information. For instance, it can show that both the violations of $a$ and $b$ occur before the violation of $c$, but it cannot precisely explain how the specification violation is triggered without capturing the full sequence of the events. Conversely, increasing $k$ to $4$ does not expose any additional information beyond what is already obtained with $k=3$ for the counterexamples in Example~\ref{ex:highK}. Therefore, if users are unsure how to choose an appropriate value of $k$, they can incrementally increase $k$ until the extracted patterns provide the level of information required for their analysis.

\myparagraph{Informativeness of our classification criterion} We summarize the information that can be delivered by our classification criterion. By Example~\ref{ex:robotTech}, we can see that, when the specification has a conjunctive structure, our criterion can extract the sufficient conditions for specification violation, which can reflect different causes of the violation. By Example~\ref{ex:robotTech} and~\ref{ex:highK}, we show that increasing $k$ can capture order information between events, which can be important in different contexts such as fault diagnosis and robot mission specifications~\cite{menghi2019specification}. 

Moreover, expressing classes using PSTL allows us to characterize them at a higher level of abstraction, focusing on semantic meanings rather than requirements with concrete valuations. For example, Class~(\ref{class:1}) in Example~\ref{ex:robotTech} captures the ordering of missed goals, regardless of when the violations occur. By instantiating the PSTL parameters with concrete values, one can further refine the analysis and filter counterexamples. For instance, by setting $t$ sufficiently large, we can exclude the less interesting cases in which $\areaone$ is missed only temporarily and instead focus on scenarios where the robot fails to visit $\areaone$ for an extended period.

\subsection{Counterexample Classification by Class Traversal}\label{sec:naiveApproach}
Based on the classification criterion presented in~\S{}\ref{sec:criterionSubsec}, we consider the counterexample classification problem stated as follows in Def.~\ref{def:ceClassProblem} and visualized in Fig.~\ref{fig:workflow}.

\begin{mydefinition}[Counterexample Classification Problems] \label{def:ceClassProblem}
Let $\varphi$ be an STL specification.  Given a finite set $\traceSet$ of counterexamples w.r.t $\varphi$ and the classification criterion given in Def.~\ref{def:classCriterion}, a counterexample classification problem requires, for each $\bw\in \traceSet$, finding all $\psi\in \SCondMinus(\varphi)$ such that $\bw\in \traceVio{\psi}$.
\end{mydefinition}

In Def.~\ref{def:ceClassProblem}, deciding whether $\bw\in \traceVio{\psi}$ involves a non-trivial search problem in the valuation space of $\psi$. 
\begin{algorithm}[!tb]
\caption{Counterexample classification by class traversal}
\label{algo:ourApproach}
\small
\begin{algorithmic}[1]
\Require An STL formula $\varphi$, a set $\traceSet$ of signals s.t. $\forall \bw\in\traceSet.\bw\not\models\varphi$, a local budget $\budget \in \Npos$
\State $S\gets \SCondMinus(\varphi)$ \label{line:initS}\Comment{obtain the set of classes}
\State $\mathsf{res}\gets \emptyset$ \label{line:resInit}\Comment{initialize $\mathsf{res}$ for result recording}
\For{$\bw \in \traceSet$} \label{line:mainLoop}
\State $\Call{Match}{S, \bw}$ \Comment{find all classes that include $\bw$}
\EndFor
\Statex
\Function{Match}{$S, \bw$}
\If{$S \neq \emptyset$}
\State select $\psi\in S$ 
\label{line:selectPsi}\Comment{select a class}
\If{\Call{MemQry}{$\psi$, \bw}}\label{line:callCheck} \Comment{membership query to $\psi$}
\State $\mathsf{res} \gets \mathsf{res} \cup \{ \langle \bw, \psi  \rangle \}$\label{line:record} \Comment{record if $\bw$ belongs to $\psi$}
\EndIf
\State $S\gets S\setminus\{\psi\}$ \label{line:updateS} \Comment{update $S$}

\State $\Call{Match}{S, \bw}$ \label{line:recursiveCall} \Comment{recursive call of \textsc{Match}}
\EndIf
\EndFunction
\Statex
\Function{MemQry}{$\psi, \bw$}
\State $\mathsf{rb_0}\gets \infty$, $i\gets 0$ \label{line:initSearch}\Comment{initialize placeholders}
    \While{$i\le \budget$ and $\mathsf{rb}_i > 0$} \label{line:enterSearch}
        \State $i\gets i + 1$
    	\State $\theta_{i} \gets \begin{cases}
            \Call{RandomSample}{} & \text{if } i\le 2  
        \\
    	    \Call{Optimizer}{ \{\langle \theta_j, \mathsf{rb}_j \rangle\}_{j\in\kset{i-1}}
            }  & \text{otherwise}\label{line:callOptimizer}
    	\end{cases} $ \Statex \Comment{suggest the next sample $\theta_i$}

    	\State $\mathsf{rb}_{i}\gets \Robust{\bw}{\psi(\theta_i)}$  \Comment{robustness as  guidance}
    	\If{$\mathsf{rb}_{i} < 0$} \Comment{found $\theta$ s.t. $\bw\not\models\psi(\theta_i)$}
        \State \Return \true
    	\EndIf
        \EndWhile
        \State\Return \false \Comment{cannot find $\theta$ s.t. $\bw\not\models\psi(\theta_i)$}
\EndFunction
\end{algorithmic}
\end{algorithm}
Our algorithm is presented in Alg.~\ref{algo:ourApproach},  
which essentially consists of a traversal of the counterexample classes in $\SCondMinus(\varphi)$. 

Alg.~\ref{algo:ourApproach} starts with loading all the classes to a set $S$ (Line~\ref{line:initS}) and initializing $\mathsf{res}$ for result recording (Line~\ref{line:resInit}). Then, for each signal $\bw$, it enters the loop of identifying the classes $\bw$ belongs to (Line~\ref{line:mainLoop}). 
At each loop, it selects a class $\psi\in S$ (Line~\ref{line:selectPsi}) and checks whether $\bw$ belongs to $\psi$ by calling \textsc{MemQry} (Line~\ref{line:callCheck}). 

In function \textsc{MemQry}, Alg.~\ref{algo:ourApproach} performs the membership query of $\bw$ to $\psi$, i.e., checking whether there exists a $\theta$ such that $\bw\not\models\psi(\theta)$ (Line~\ref{line:enterSearch}). Essentially, this requires solving a search problem as follows:
\begin{align*}
    \min \quad \Robust{\bw}{\psi(\theta)} \qquad
s.t. \quad \theta\in\valueSpace
\end{align*}
The search goal involves finding a parameter valuation $\theta$ such that $\Robust{\bw}{\psi(\theta)} < 0$, i.e., $\bw\not\models\psi(\theta)$, which signifies that $\bw\in\traceVio{\psi}$, and therefore, it is not necessary to find the optimal $\theta$ but it suffices to terminate as soon as $\Robust{\bw}{\psi(\theta)} < 0$. As shown in Line~\ref{line:callOptimizer}, this is performed by calling an off-the-shelf optimization solver, that suggests a new valuation $\theta_i$ at each time with the aim of minimizing $\Robust{\bw}{\psi(\theta_i)}$. Typically, at an initial stage (e.g., $i\le 2$), these solvers may randomly sample candidate $\theta$ from the search space; later, based on the sampling history, the optimizer can suggest new $\theta$ according to some heuristics. Note that here as the search space $\valueSpace$ consists of infinitely many valuations and the objective function $\Robust{\bw}{\psi(\theta_i)}$ can be highly non-linear, it is difficult to provably show the existence of such $\theta$ that makes $\bw\models\psi(\theta)$. Nevertheless, as the computation of each objective function value involves monitoring of a signal against an STL formula $\psi(\theta)$ that is not expensive, modern optimization solvers can still handle this problem effectively. 

If such $\theta$ is found, Alg.~\ref{algo:ourApproach} records $\langle\bw, \psi\rangle$, signifying that $\bw$ belongs to the class $\psi$ (Line~\ref{line:record}). Then
it updates $S$ by removing $\psi$ from it (Line~\ref{line:updateS}) and recursively calls \textsc{Match} to check whether $\bw$ belongs to the next class (Line~\ref{line:recursiveCall}).

Essentially, Alg.~\ref{algo:ourApproach} enumerates all classes $\psi\in \SCondMinus(\varphi)$ for each $\bw\in \traceSet$ to check whether each $\bw$ is included in each $\psi$. Therefore, the complexity of Alg.~\ref{algo:ourApproach} is $O(|\numClass|\cdot \budget\cdot |\traceSet|)$, where $|\numClass|$ is the number of classes. Note that $|\numClass|$ is exponential to the selected $k$ (see Def.~\ref{def:classCriterion}), so Alg.~\ref{algo:ourApproach} is relatively expensive and does not scale well with the growth of $k$.

\section{An Efficient Classification Approach}\label{sec:efficient}
While the approach in~\S{}\ref{sec:naiveApproach} can solve the classification problem, it may not be efficient, because the number of classes grows exponentially with the value of $k$ in Def.~\ref{def:classCriterion} for temporal operators. To mitigate this issue, we first present our insight about the relation between different classes, which can be formalized as a partial order over the classes; we propose two approaches leveraging this insight, which can accelerate classification by skipping many unnecessary class queries.

\subsection{Partial Order over Counterexample Classes}\label{sec:partialOrder}
We define a partial order over different classes in $\SCondPlus(\varphi)$ and $\SCondMinus(\varphi)$, based on the inclusion relation of their denoted sets of signals.

\begin{mydefinition}[Inclusion Relation]\label{def:inclusion}
Given an STL formula $\varphi$ and given the settings in Def.~\ref{def:classCriterion}, 
 we define an order $\smallerSatEq$ over two different classes $\psi_1, \psi_2\in \SCondPlus(\varphi)$, as follows, 
 \begin{math}
     \psi_1\smallerSatEq\psi_2 \text{ if and only if } \traceSat{\psi_1}\subseteq\traceSat{\psi_2}
 \end{math};
 similarly, we define an order $\smallerVioEq$ over two different classes $\psi_1, \psi_2\in \SCondMinus(\varphi)$, as follows:
 \begin{math}
     \psi_1\smallerVioEq \psi_2 \text{ if and only if } \traceVio{\psi_1} \subseteq\traceVio{\psi_2}
 \end{math}.
 Alternatively, we can treat $\smallerSatEq(\varphi)$ and $\smallerVioEq(\varphi)$ as sets of pairs of the classes of $\varphi$, and write $\pair{\psi_1}{\psi_2}\in \smallerSatEq(\varphi)$ to denote $\psi_1\smallerSatEq\psi_2$, and write $\pair{\psi_1}{\psi_2}\in \smallerVioEq(\varphi)$ to denote $\psi_1\smallerVioEq\psi_2$. 
\end{mydefinition}

Def.~\ref{def:inclusion} describes the properties expected to hold by the order, but it does not give a constructive algorithm on how to identify such an order for a given class set. 
In Def.~\ref{def:orderIdentify}, we present an approach for identifying the orders $\orderSat(\varphi)$ and $\orderVio(\varphi)$ respectively in $\SCondPlus(\varphi)$ and $\SCondMinus(\varphi)$. Later by Theorem~\ref{lem:identification} and Theorem~\ref{theo:relationComplete}, we show that Def.~\ref{def:orderIdentify} precisely captures the desired semantic of the inclusion relation specified in Def.~\ref{def:inclusion}.

\begin{table*}[!tb]
    \caption{A constructive approach to identify $\orderSat(\varphi)$ and $\orderVio(\varphi)$}
    \label{tab:order}
    \centering
        \begin{align*} 
        \midrule
        &\orderSat(\alpha) \;\Defeq\; \{\pair{\bot}{\alpha}, \pair{\alpha}{\alpha}, \pair{\bot}{\bot}\} 
        \qquad \orderSat(\neg\varphi) \;\Defeq\; \left\{\pair{\neg\psi}{\neg\sigma} \setVert \pair{\psi}{\sigma}\in \orderVio(\varphi)\right\}
           \\
        &\orderSat(\varphi_1\land \varphi_2) \;\Defeq\; \left\{ \pair{\psi_1\land\psi_2}{\sigma_1\land\sigma_2}\setVert \pair{\psi_1}{\sigma_1}\in \orderSat(\varphi_1), \pair{\psi_2}{\sigma_2}\in \orderSat(\varphi_2) \right\} \\
        & \orderSat(\varphi_1\lor \varphi_2) \;\Defeq\; \left\{ \pair{\psi_1\lor\psi_2}{\sigma_1\lor\sigma_2}
         \setVert 
\pair{\psi_1}{\sigma_1}\in \orderSat(\varphi_1), \pair{\psi_2}{\sigma_2}\in \orderSat(\varphi_2) 
        \right\}    
               \\
        & \textstyle\orderSat(\Box_I\varphi) \;\Defeq\; \left\{ \pair{\bigwedge_{i\in\kset{k}} \Box_{[l_i, u_i]}\psi_i}{\bigwedge_{i\in\kset{k}} \Box_{[l_i, u_i]}\sigma_i} \setVert \forall i\in\kset{k}.\;\pair{\psi_i}{\sigma_i} \in \orderSat(\varphi) \right\}  \\
        & \textstyle\orderSat(\Diamond_I\varphi) \;\Defeq\; \left\{ \pair{\bigvee_{i\in\kset{k}}\Diamond_{[l_i, u_i]}\psi_i}{\bigvee_{i\in\kset{k}}\Diamond_{[l_i,u_i]}\sigma_i}
            \setVert
            \forall i \in\kset{k}.\;\pair{\psi_i}{\sigma_i} \in \orderSat(\varphi)
        \right\} \\
    &\orderSat(\varphi_1\UntilOp{I}\varphi_2) \;\Defeq\;  \left\{ 
    \left\langle
    \begin{array}{l}
         \exists i \in \kset{k}
      \Bigl(
        \psi_i \UntilOp{[l_i,u_i]} \sigma_i
        \land
        \forall j<i,\,
        \Box_{[l_{j},u_{j}]} \xi_{j}
      \Bigr) \\
         \exists i \in \kset{k}
      \Bigl(
        \psi'_i \UntilOp{[l_i,u_i]} \sigma'_i
        \land
        \forall j<i,\,
        \Box_{[l_{j},u_{j}]} \xi'_{j}
      \Bigr) 
    \end{array} \right\rangle \setVert
    \begin{array}{ll}
        \forall i\in\kset{k}, 
        &\pair{\psi_i}{\psi'_i} \in \orderSat(\varphi_1), 
        \pair{\sigma_i}{\sigma'_i}\in \orderSat(\varphi_2) \\
        \forall j < i,
         &\pair{\xi_j}{\xi'_j}\in \orderSat(\varphi_1)
    \end{array}
    \right\}
\\ \midrule
        &\orderVio(\alpha) \;\Defeq\; \{\pair{\top}{\alpha}, \pair{\alpha}{\alpha}, \pair{\top}{\top}\} \qquad
        \orderVio(\neg\varphi) \;\Defeq\; \left\{\pair{\neg\psi}{\neg\sigma} \mid \pair{\psi}{\sigma}\in \orderSat(\varphi)\right\}
           \\
        &\orderVio(\varphi_1\land \varphi_2) \;\Defeq\; \left\{ 
        \pair{\psi_1\land\psi_2}{\sigma_1\land\sigma_2}
        \setVert
        \pair{\psi_1}{\sigma_1} \in \orderVio(\varphi_1), \pair{\psi_2}{\sigma_2}  \in \orderVio(\varphi_2)
        \right\}    
        \\
        & \orderVio(\varphi_1\lor \varphi_2) \;\Defeq\;  \left\{ \pair{\psi_1\lor\psi_2}{\sigma_1\lor\sigma_2} \setVert \pair{\psi_1}{\sigma_1}\in \orderVio(\varphi_1), \pair{\psi_2}{\sigma_2}\in \orderVio(\varphi_2) \right\}
               \\
        & \textstyle\orderVio(\Box_I\varphi) \;\Defeq\; \left\{ \pair{\bigwedge_{i\in \kset{k}} \Box_{[l_i, u_i]}\psi_i}{\bigwedge_{i\in\kset{k}} \Box_{[l_i, u_i]}\sigma_i} \setVert \forall i\in\kset{k}.\;\pair{\psi_i}{\sigma_i} \in \orderVio(\varphi) \right\} 
        \\
        &\textstyle\orderVio(\Diamond_I\varphi) \;\Defeq\; \left\{ \pair{\bigvee_{i\in\kset{k}}\Diamond_{[l_i, u_i]}\psi_i}{\bigvee_{i\in\kset{k}}\Diamond_{[l_i,u_i]}\sigma_i}
            \setVert
            \forall i \in\kset{k}.\;\pair{\psi_i}{\sigma_i} \in \orderVio(\varphi)
        \right\}  \\
        &\orderVio(\varphi_1\UntilOp{I}\varphi_2) \;\Defeq\; 
        \left\{ 
    \left\langle
    \begin{array}{l}
         \exists i \in \kset{k}
      \Bigl(
        \psi_i \UntilOp{[l_i,u_i]} \sigma_i
        \land
        \forall j<i,\,
        \Box_{[l_{j},u_{j}]} \xi_{j}
      \Bigr) \\
         \exists i \in \kset{k}
      \Bigl(
        \psi'_i \UntilOp{[l_i,u_i]} \sigma'_i
        \land
        \forall j<i,\,
        \Box_{[l_{j},u_{j}]} \xi'_{j}
      \Bigr) 
    \end{array} \right\rangle \setVert
    \begin{array}{ll}
        \forall i\in\kset{k}, &
        \pair{\psi_i}{\psi'_i} \in \orderVio(\varphi_1), 
        \pair{\sigma_i}{\sigma'_i}\in \orderVio(\varphi_2) \\
        \forall j < i, &
         \pair{\xi_j}{\xi'_j}\in \orderVio(\varphi_1)
    \end{array}
    \right\} \\ \midrule
        \end{align*} 
\end{table*}
\begin{mydefinition}[Order Construction]\label{def:orderIdentify}
Let $\varphi$ be an STL formula, and assume the same settings in Def.~\ref{def:classCriterion}. We identify the relation $\orderSat(\varphi)$ as a set of pairs $\pair{\psi}{\sigma}$ of classes as defined in Table~\ref{tab:order}, where $\psi\in \SCondPlus(\varphi)$ and $\sigma\in\SCondPlus(\varphi)$.
We also identify the relation $\orderVio(\varphi)$ as a set of pairs $\pair{\psi}{\sigma}$ of classes, as defined in Table~\ref{tab:order}, where $\psi\in \SCondMinus(\varphi)$ and $\sigma\in\SCondMinus(\varphi)$.
\end{mydefinition}

\begin{mylemma}\label{lem:consistency}
   Given the definition of $\smallerSatEq(\varphi)$ and $\smallerVioEq(\varphi)$ in Def.~\ref{def:orderIdentify}, for any pair $\pair{\psi}{\sigma}\in \smallerSatEq(\varphi)$, it holds that $\psi \in \SCondPlus(\varphi)$ and $\sigma\in\SCondPlus(\varphi)$, and for any pair $\pair{\psi}{\sigma}\in \smallerVioEq(\varphi)$, it holds that $\psi \in \SCondMinus(\varphi)$ and $\sigma\in\SCondMinus(\varphi)$.
\end{mylemma}

Lemma~\ref{lem:consistency} validates that each object in each pair of the relations $\orderSat(\varphi)$ and $\orderVio(\varphi)$ must be a class from $\SCondPlus(\varphi)$ or $\SCondMinus(\varphi)$ in Def.~\ref{def:classCriterion}. Furthermore, we formally state the soundness of Def.~\ref{def:orderIdentify}, i.e., the correctness of the order between different classes, in Theorem~\ref{lem:identification} and the completeness, i.e., Def.~\ref{def:orderIdentify} covers all inclusion relations over the classes of a given $\varphi$, in Theorem~\ref{theo:relationComplete}. The proofs are left in Appendix~\ref{app:theo3} and~\ref{app:theo4}.

\begin{mytheorem}[Soundness of $\orderSat(\varphi)$]\label{lem:identification}
     Given an STL formula $\varphi$, if $ \pair{\psi}{\sigma} \in\smallerSatEq(\varphi)$, it holds that $\traceSat{\psi}\subseteq \traceSat{\sigma}$; similarly,
     if $\pair{\psi}{\sigma} \in\smallerVioEq(\varphi)$, it holds that $\traceVio{\psi}\subseteq \traceVio{\sigma}$.
\end{mytheorem}

\begin{mycorollary}\label{lem:partial}
    The $\smallerSatEq$ and $\smallerVioEq$ in Def.~\ref{def:orderIdentify} are partial orders.
\end{mycorollary}

\noindent{\it Proof sketch.}
    This corollary holds because the relations in Def.~\ref{def:orderIdentify} are reflexive, antisymmetric, and transitive.
\qed

\begin{mytheorem}[Completeness of $\orderSat(\varphi)$]\label{theo:relationComplete}
    For any classes $\psi, \sigma \in \SCondPlus(\varphi)$ of a given STL formula $\varphi$, if it holds $\traceSat{\psi}\subseteq \traceSat{\sigma}$, it implies that $ \pair{\psi}{\sigma} \in\smallerSatEq(\varphi)$; if it holds  $\traceVio{\psi}\subseteq \traceVio{\sigma}$, it implies that $\pair{\psi}{\sigma} \in\smallerVioEq(\varphi)$.
\end{mytheorem}

\smallskip
Notably, the relations $\smallerSatEq$ and $\smallerVioEq$ are both partial orders, as stated in Corollary~\ref{lem:partial}. Intuitively, as they resemble the inclusion relation over the satisfiable/violable sets of the classes, similar to inclusion relations of sets, they are also partial orders.


\subsection{Counterexample Classification by Binary Search}\label{sec:bsAlg}
As stated in Corollary~\ref{lem:partial}, the $\smallerVioEq$ relation in Def.~\ref{def:orderIdentify} makes  the set $\SCondMinus(\varphi)$ of classes a partial order set w.r.t. $\smallerVioEq$, so the classes in $\SCondMinus(\varphi)$ can be  constructed as a directed acyclic graph $\graph = \langle \nodes, \edges \rangle$, as follows:
\begin{compactitem}
    \item Each node $\node\in \nodes$ represents a class $\psi\in \SCondMinus(\varphi)$;
    \item An edge $\edge\in\edges$ exists between two nodes $\psi$ and $\sigma$ only if $\sigma$ immediately succeeds $\psi$, i.e., it holds that $\psi\smallerVioEq \sigma$ and  there exists no $\iota\in \SCondMinus(\varphi)$ such that $\psi\smallerVioEq\iota$ and $\iota\smallerVioEq \sigma$.
\end{compactitem} 
We define a path in $\graph$ as a sequence $\node_1\ldots \node_n$, such that each $\node_{i+1}$ immediately succeeds $\node_i$  ($i\in\kset{n-1}$), i.e., there exists an edge between  $\node_i$ and $\node_{i+1}$. Notably, each path is \emph{monotonic}, in the sense that, given a signal $\bw$, if $\bw\in\traceVio{\node_{i}}$, it holds that $\bw\in \traceVio{\node_{i+1}}$; reversely, if $\bw\not\in\traceVio{\node_{i+1}}$, it holds that $\bw\not\in \traceVio{\node_i}$. This can be formally derived by Corollary~\ref{cor:order} (here Corollary~\ref{cor:order} discusses the violation cases only, because later we use it for classification of counterexamples), and it can be proved based on Theorem~\ref{lem:identification} by contrapositive.

\begin{mycorollary}\label{cor:order}
    Given $\pair{\psi}{\sigma}\in \smallerVioEq(\varphi)$, if $\bw \in \traceVio{\psi}$, it holds that $\bw\in \traceVio{\sigma}$; if $\bw\not\in\traceVio{\sigma}$, it holds that $\bw\not\in\traceVio{\psi}$. 
\end{mycorollary}

In other words, for a given counterexample $\bw$, in each path, there exists a boundary that separates the classes that include $\bw$ and that do not include $\bw$. Specifically, it is possible that all the classes in a path include or not include $\bw$. To this end, the classification problem for a given counterexample can be reduced to finding the boundary in each path of the graph.

\medskip
Moreover, the monotonicity of each path allows us to apply binary search to identify the boundary in each path. Whenever we identify that $\bw\in \traceVio{\psi}$ or $\bw\not\in\traceVio{\psi}$, we can skip quantities of classes that are unnecessary to check, due to the monotonicity of the path. In this way, we can save many non-trivial class membership queries, thereby improving the efficiency of classification. Below, we present two approaches that leverage this insight to accelerate classification.

\begin{algorithm}[!tb]
\caption{\alwmid: Always pruning at the midpoint}
\label{algo:alwmid}
\small
\begin{algorithmic}[1]
\Require An STL formula $\varphi$, a set $\traceSet$ of signals s.t. $\forall \bw\in\traceSet.\bw\not\models\varphi$, the DAG \graph constituted by classes in $\SCondMinus(\varphi)$
\Function{Match}{$S, \bw$}
\If{$S \neq \emptyset$}
\State select the longest path $\psi^1\ldots \psi^l$ in graph $\graph$ \label{line:selectPathBS}
\State $m \gets \lceil \frac{1+l}{2} \rceil$ \label{line:midBS}
\If{\Call{MemQry}{$\psi^m, \bw$}} \label{line:memberBS}
\For{$\psi^*\in \{\psi\in S \mid \psi^m \smallerVioEq \psi\}$}
\State $\mathsf{res} \gets \mathsf{res} \cup \{ \langle \bw, \psi^*  \rangle \}$\label{line:recordBS}
\State $S\gets S\setminus \{\psi^*\}$ \label{line:includeSkipBS}
\EndFor
\Else
\For{$\psi^* \in \{\psi\in S\mid \psi\smallerVioEq\psi^m \}$}
\State $S\gets S\setminus \{\psi^*\}$\label{line:notIncludeSkip}
\EndFor
\EndIf
\State $\Call{Match}{S, \bw}$
\EndIf
\EndFunction
\end{algorithmic}
\end{algorithm}
\myparagraph{\alwmid: Always pruning at the midpoint} The first algorithm, \alwmid, is designed as in Alg.~\ref{algo:alwmid}. In order to prune as many classes as possible, it iteratively selects the longest path and performs membership query with the class at the midpoint of the path; by doing that, it can prune half of the classes on the path. We only present the different part w.r.t. Alg.~\ref{algo:ourApproach}, i.e., the function \textsc{Match}, regarding how we explore the classes. 

Alg.~\ref{algo:alwmid} starts with selecting the longest path $\psi^1\ldots \psi^l$ of length $l$ ($l\in \Npos$) in the graph $\graph$ (Line~\ref{line:selectPathBS}), by using a graph search algorithm. 
Given the selected path, Alg.~\ref{algo:alwmid} performs a membership query for the class $\psi^m$ in the middle of the path (Line~\ref{line:midBS}-\ref{line:memberBS}). If $\bw$ is included in the class $\psi^m$, all classes $\psi$ such that $\psi^m\smallerVioEq \psi$ can be considered to include $\bw$ (Line~\ref{line:recordBS}), so they are not needed to be checked and removed from $S$ (Line~\ref{line:includeSkipBS}); if $\bw$ is not included in the class $\psi^m$, all classes $\psi$ such that $\psi\smallerVioEq \psi^m$ can be considered not to include $\bw$, so they are also removed from $S$ (Line~\ref{line:notIncludeSkip}). In this way, after every membership query, we can skip all the classes that are greater or less, in terms of $\smallerVioEq$, than the class being checked, thereby accelerating classification significantly.

Compared to Alg.~\ref{algo:ourApproach}, Alg.~\ref{algo:alwmid} scales better for large $k$, because every query with a class can prune half of the classes on a selected path, so the number of queries can be reduced from $O(|\numClass|)$ to $O(\log(|\numClass|))$, given the inclusion relation in Def.~\ref{def:orderIdentify}.  On the other hand, it introduces additional overhead for each query, namely, the search for the longest path in $\graph$, which can take $O(|\numClass|)$. Therefore, the complexity of Alg.~\ref{algo:alwmid} is $O(|\traceSet|\cdot\log(|\numClass|)\cdot(|\numClass|+\budget))$; since the search of longest path exhibits a decreasing complexity as the algorithm progresses, the actual performance of Alg.~\ref{algo:alwmid} can be much better than Alg.~\ref{algo:ourApproach}.

\begin{algorithm}[!tb]
\caption{\longbs: Binary search for longest paths}
\label{algo:longbs}
\small
\begin{algorithmic}[1]
\Require An STL formula $\varphi$, a set $\traceSet$ of signals s.t. $\forall \bw\in\traceSet.\bw\not\models\varphi$, the DAG \graph constituted by classes in $\SCondMinus(\varphi)$
\Function{Match}{$S, \bw$}
\If{$S \neq \emptyset$}
\State select the longest path $\psi^1\ldots \psi^l$ in graph $\graph$ \label{line:selectPathLBS}
\State $\mathit{start}\gets 1$, $\mathit{end}\gets l$,
$m \gets \lceil \frac{1+l}{2} \rceil$
\While{$\mathit{start} \le \mathit{end}$} \label{line:BSLoopLBS}
    \If{\Call{MemQry}{$\psi^m, \bw$}} \label{line:memberLBS}
        \For{$\psi^*\in \{\psi\in S \mid \psi^m \smallerVioEq \psi\}$}
            \State $\mathsf{res} \gets \mathsf{res} \cup \{ \langle \bw, \psi^*  \rangle \}$\label{line:recordLBS}
            \State $S\gets S\setminus \{\psi^*\}$ \label{line:includeSkipLBS}
        \EndFor
        \State $\mathit{end}\gets m - 1$
    \Else
        \For{$\psi^* \in \{\psi\in S\mid \psi\smallerVioEq\psi^m \}$}
        \State $S\gets S\setminus \{\psi^*\}$\label{line:notIncludeSkipLBS}
        \EndFor
        \State $\mathit{start}\gets m + 1$
    \EndIf
    \State $m \gets \lceil \frac{1+l}{2} \rceil$ \label{line:midLBS}
\EndWhile
\State $\Call{Match}{S, \bw}$ \label{line:callMatchLBS}
\EndIf
\EndFunction
\end{algorithmic}
\end{algorithm}
\myparagraph{\longbs: Binary Search for longest paths} While Alg.~\ref{algo:alwmid} saves a significant number of membership queries, the path selection at Line~\ref{line:selectPathBS}, may introduce non-negligible overhead, especially when the graph $\graph$ is large. To address this issue, we propose another algorithm, \longbs, presented in Alg.~\ref{algo:longbs}, which reduces the number of path selections while still leveraging the ordering to prune membership queries. 

Alg.~\ref{algo:longbs} also selects the longest path in the graph $\graph$ (Line~\ref{line:selectPathLBS}), as Alg.~\ref{algo:alwmid} does; after that, it enters a binary search loop (Line~\ref{line:BSLoopLBS}), in which it iteratively selects the midpoint of the path segment including the undecided classes, and prunes the classes in $S$ according to the query result (Line~\ref{line:memberLBS}-\ref{line:midLBS}), similarly to Alg.~\ref{algo:alwmid}. Unlike Alg.~\ref{algo:alwmid}, it invokes the recursive call of \textsc{Match} (Line~\ref{line:callMatchLBS}) only after finishing a binary search loop, so it can avoid  doing the expensive path selection frequently.

For each signal, it selects a longest path once and performs a full binary search over it with $O(\log l)$ queries, before moving to the next path. With $\numPath$ paths selected in total, the overall cost per signal is $O(\numPath \cdot |\numClass|)$ for path searches and $O(\numPath \cdot \log l \cdot \budget)$ for membership queries. Here, it is difficult to characterize $\numPath, l$ by $|\numClass|$; in the worst case if $O(\log l)$ is $O(1)$, then $\numPath$ becomes $O(|\numClass|)$ and the worst-case complexity becomes $O(|\traceSet|\cdot|\numClass|\cdot(|\numClass|+ \budget))$. However, this means that the number of paths selected equals to the number of classes, which is not likely the real case. Similar to Alg.~\ref{algo:alwmid}, the pruning can make $\numPath$ much smaller so the actual performance can be much better.

\section{Experimental Evaluation}\label{sec:evaluation}
We present the experimental evaluation of our proposed framework. All code and data are available online: {\url{https://github.com/parvkpr/CE-class-python}}. 

\subsection{Experiment Settings}\label{sec:experimentSetting}

\myparagraph{Benchmarks}
We adopt three systems for evaluation, including two Simulink systems widely adopted as benchmarks in the hybrid system community~\cite{ARCHCOMP25Falsification} and the robotic system, adapted from~\cite{kapoor2024safe}, described in Example~\ref{ex:robotIntro}.  
\begin{compactitem}
    \item {\it Automatic transmission (AT)} originates from \textsc{MathWorks}, and has been used as a standard benchmark in ARCH competition~\cite{ARCHCOMP18Falsification, ARCHCOMP25Falsification}. It has multiple input/output signals, including $\gear$, $\speed$ and $\rpm$, etc. 

\spec{AT}{1} is a simple safety requirement, which requires that the \speed of the car never exceeds 100. 
 \begin{align*}
\spec{AT}{1} \equiv \Box_{[0, 30]}(\speed < 100)
\end{align*}

\spec{AT}{2} is a conjunctive requirement, meaning that, during $[0,30]$, \speed should always be less than $90$ and \rpm should be always less than $4000$. 
\begin{align*}
\spec{AT}{2} \equiv \Box_{[0, 30]}(\speed < 90 \land \rpm < 4000)
\end{align*}

\spec{AT}{3} is a liveness property, requiring that, there exists a moment at which both \speed is greater than 70 and \rpm is greater than 3800. 
\begin{align*}
 \spec{AT}{3} \equiv \Diamond_{[0,30]}(\speed > 70 \land \rpm > 3800)
\end{align*}

\spec{AT}{4} aims to test whether the system can react to a braking action timely. When \brake is lower than 250, it requires that the \speed and the \rpm should decrease to the expected level in 5 secs. 
\begin{align*}
    \spec{AT}{4} &\equiv \Box_{[0,30]}(\brake < 250 \to \Diamond_{[0,5]}(\reaction)) \\
    \reaction &\equiv (\speed < 30 \land \rpm < 2000)
\end{align*}

Moreover, we consider the following  \spec{AT}{\fals} for the purpose of falsification. \spec{AT}{\fals} encodes multiple goals expected by one system execution. While the initial \speed is high, the first goal is to lower it down; once it is achieved, the second and the third goal should be achieved in 5 secs, the former aiming to raise \rpm, the latter leading car to a stable state where $\speed < 110$ periodically happen. 
\begin{align*}
    \spec{AT}{\fals} &\equiv \Diamond_{[10,15]}(\targetone \land \Diamond_{[0,5]}(\targettwo \land \stable))  \\
    \targetone & \equiv\speed < 85, \quad
    \targettwo \equiv \rpm > 800 \\
    \stable &\equiv\Box_{[0,10]}\Diamond_{[0,5]}(\speed < 110)
\end{align*}

\item {\it Abstract Fuel Control (AFC)} is a powertrain control system released by Toyota~\cite{jin2014powertrain}. It implements a complicated control logic, and the goal is to ensure that the \emph{air-to-fuel} ratio $\af$, does not deviate too much from $\afref$.

\spec{AFC}{1} which requires that, during $[0,40]$, there should exist a moment, since which, the stable state of \af, neither too high nor too low, should hold for at least 10 seconds. 
\begin{align*}
 \spec{AFC}{1} \equiv \Diamond_{[0,40]}\Box_{[0,10]}(\af - \afref \in (-0.05, 0.05))
\end{align*}

\item {\it Robotic System (Rob)} is the example described in Example~\ref{ex:robotIntro}. We consider a concrete \spec{Rob}{} as follows: 
\begin{align*}
    & I_1 = [0,25], \quad I_2 = [0,25],\quad I = [0,50] \\
    &\areaone \in [1,5] \times [1,5],\quad  
    \areatwo \in [15,19] \times [15,19] \\
    &\danger \in [8,12] \times [8,12]
\end{align*}

\end{compactitem}

\myparagraph{Hardware \& software dependencies}
Our experiments of counterexample classification are conducted on an NVIDIA RTX 5090 GPU. For classification, we use STLCG++~\cite{11077999} for GPU-based acceleration of parallel robustness evaluation and parameter synthesis. For falsification, we run on a Macbook Pro with a chip of Apple M4 Pro, 24GB memory; we use Breach~\cite{donze2010breach} as our falsification tool, which is a common selection that integrates different optimization algorithms.

\begin{figure*}[!t]
\centering
\includegraphics[width=0.8\linewidth]{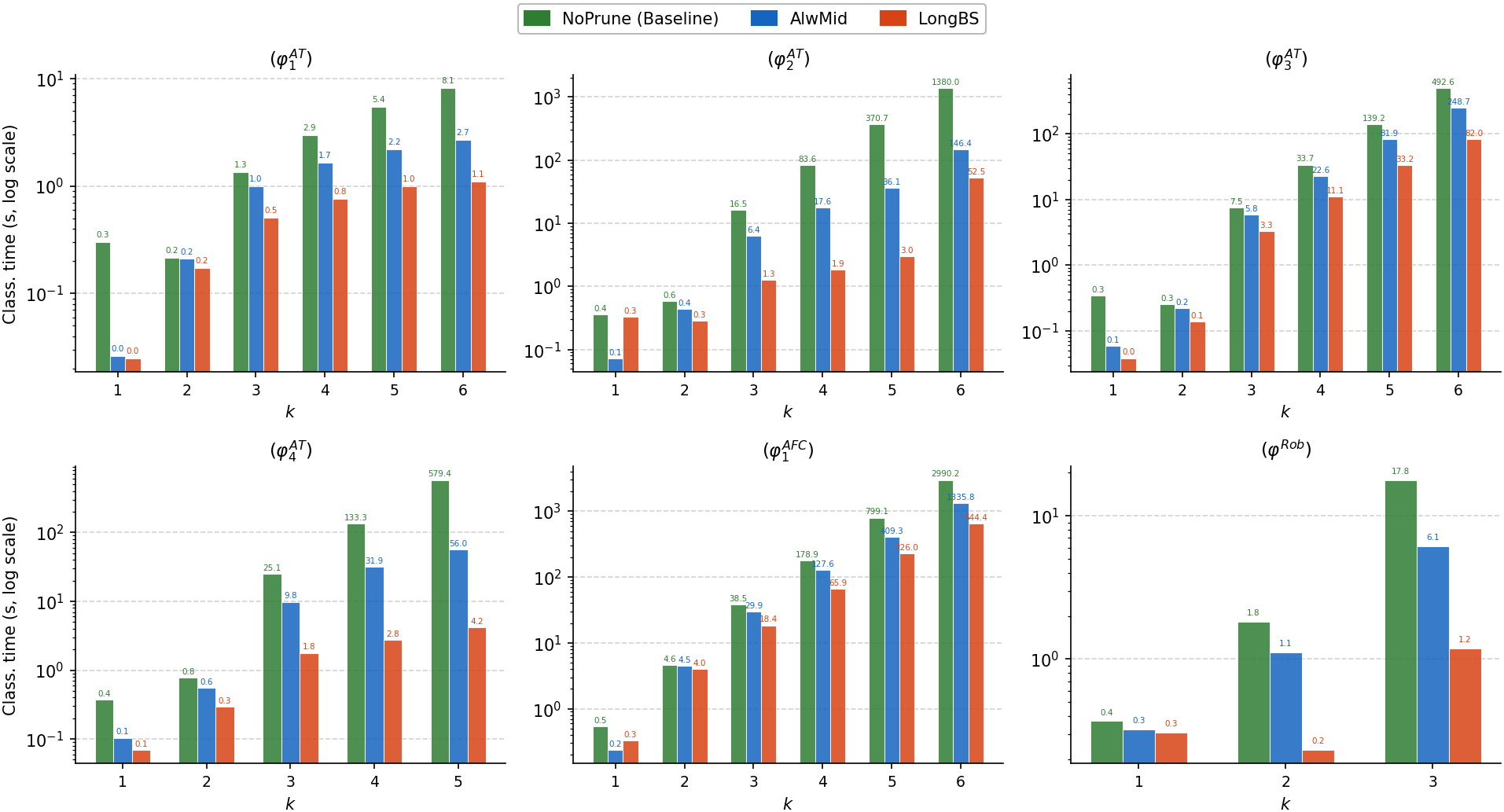}
\caption{Time costs for classifying the collected counterexamples (in secs, $y$-axis in log scale)}
\label{fig:efficiency}
\end{figure*}
\subsection{Efficiency}\label{sec:efficiency}
To evaluate efficiency, for each specification $\varphi$ and each set $W$ of counterexamples that violate $\varphi$, we apply our approaches to classify the counterexamples in $W$ into classes of $\SCondMinus(\varphi)$. The counterexample set $W$ is collected by running simulations with randomly generated input signals. For each $\varphi$, we collect 100 counterexamples as $W$ and record the time costs (in secs) for classifying them.  We also assess our performance under varying $k$, by setting $k\in \{1,2,3,4,5,6\}$ for AT and AFC, and $k\in\{1,2,3\}$ for Rob (all for the outmost temporal operator; as that interval of \spec{Rob}{} is relatively short, we try out less $k$s than others). We set 3000 secs as the timeout for each classification. The results are presented in Fig.~\ref{fig:efficiency}.

\myparagraph{Comparison with baseline} We take the straightforward classification approach presented in Alg.~\ref{algo:ourApproach} as a baseline and assess whether  \alwmid and \longbs can be more efficient. We observe that both of the approaches, \alwmid and \longbs, outperform the baseline across different selections of $k$, thanks to the proposed acceleration strategies in~\S{}\ref{sec:bsAlg}. By leveraging the partial order over the classes, we can skip quantities of unnecessary membership queries, which is helpful to improve efficiency. This is particularly evident when $k$ is relatively large. For example, while our performance for \spec{AT}{2} is marginally better than the baseline when $k = 1$ and $k = 2$, both our approaches achieve significant improvement when $k = 4, 5$ and $6$, which demonstrates the effectiveness of our approaches.

\myparagraph{\alwmid vs. \longbs}
We observe that, in general, \longbs outperforms \alwmid, especially when $k$ becomes large. When $k = 1$ or $k = 2$, these two approaches are often comparable, but when $k = 3$ to $k = 6$, \longbs shows evident performance advantages over \alwmid. This can be explained by the difference between the two approaches, namely, while \alwmid keeps searching for the longest path in the graph to perform membership queries, \longbs takes a different strategy that only searches for the next path after it finishes assessing a selected path. This is particularly useful when the graph is large, in which case searching for paths can take non-negligible overhead. Our experimental results indeed demonstrate the importance of the strategy taken by \longbs.  
 
\myparagraph{Scalability} 
Our results also demonstrate the scalability of our approaches. The baseline can scale to $k = 6$ for simple specifications such as \spec{AT}{1}, \spec{AT}{2} and \spec{AT}{3}, but struggles for others. As expected, for each specification, the time cost increases exponentially as $k$ increases, due to the exponential increase of the number of classes. In comparison, \alwmid and \longbs scale better than the baseline, and can terminate within a reasonable time for most of the specifications. In particular, when $k = 6$ for \spec{AT}{2}, \alwmid can reduce 89.4\%, while \longbs can reduce 96.2\% of the cost of the baseline. For \spec{AT}{4}, while both approaches fail to terminate within the time budget for $k = 6$, they can terminate for $k = 5$, which is sufficient for many real-world scenarios.

\subsection{Effectiveness of Our Approach}\label{sec:effect}

\myparagraph{Preferential falsification}
We show how our approach is helpful to falsification, i.e., the technique used to find counterexamples to refute a given specification. As mentioned, existing falsification mostly aims to show existence of counterexamples, so by design, they terminate when they find such a system output that violates the specification. By contrast, we can firstly use our classification criterion to generate different classes, each of which identifies a different \emph{sufficient condition} for violation and highlights a distinct violation pattern, and then employ the generated class to guide falsification to produce diverse counterexamples. 

Notably, in literature, there is a line of work~\cite{sato2021constrained, mathesen2021efficient} that specifically considers falsification against conjunctive requirements. This can be motivated for different reasons, e.g., in~\cite{sato2021constrained}, they address the industrial needs for counterexamples in specific regions in state space. Due to the conjunctive structure, in those works the specifications can be easily decomposed into their different clauses, which correspond to different sufficient conditions for violation of the original specifications; however, for general specifications, it may be difficult to figure out such different sufficient conditions by hand. In that case, by our approach, we can efficiently obtain them, and then use them to guide falsification.

We select \spec{AT}{\fals} to showcase our approach, which has a relatively complex structure and multiple Boolean connectives and temporal operators nested with each other.
By applying our criterion (with $k=1$ for each temporal operator), we obtain 8 classes and we focus on the following three classes, each of which represents a violation due to an atomic reason:
\begin{align*}
    \psi_1 &\equiv \Diamond_{[10,15]}(\speed < 85) \\
    \psi_2 &\equiv \Diamond_{[10,15]}(\Diamond_{[0,5]}(\rpm > 800))\\
    \psi_3 &\equiv \Diamond_{[10, 15]}(\Diamond_{[0,5]}(\Box_{[0,10]}(\Diamond_{[0,5]}(\speed < 110))))
\end{align*}

To demonstrate this approach, we compare it with the normal falsification guided by the original specification \spec{AT}{\fals}. Our settings are as follows. For each falsification trial, we set 200 secs as time budget; to account for algorithmic stochasticity, we repeat falsification trials for \spec{AT}{\fals} for 60 times, and repeat falsification for each of $\psi_1$, $\psi_2$ and $\psi_3$ for 20 times. We assess how many counterexamples each approach can find regarding different classes.

\begin{figure}
    \centering
    \includegraphics[width=0.6\linewidth]{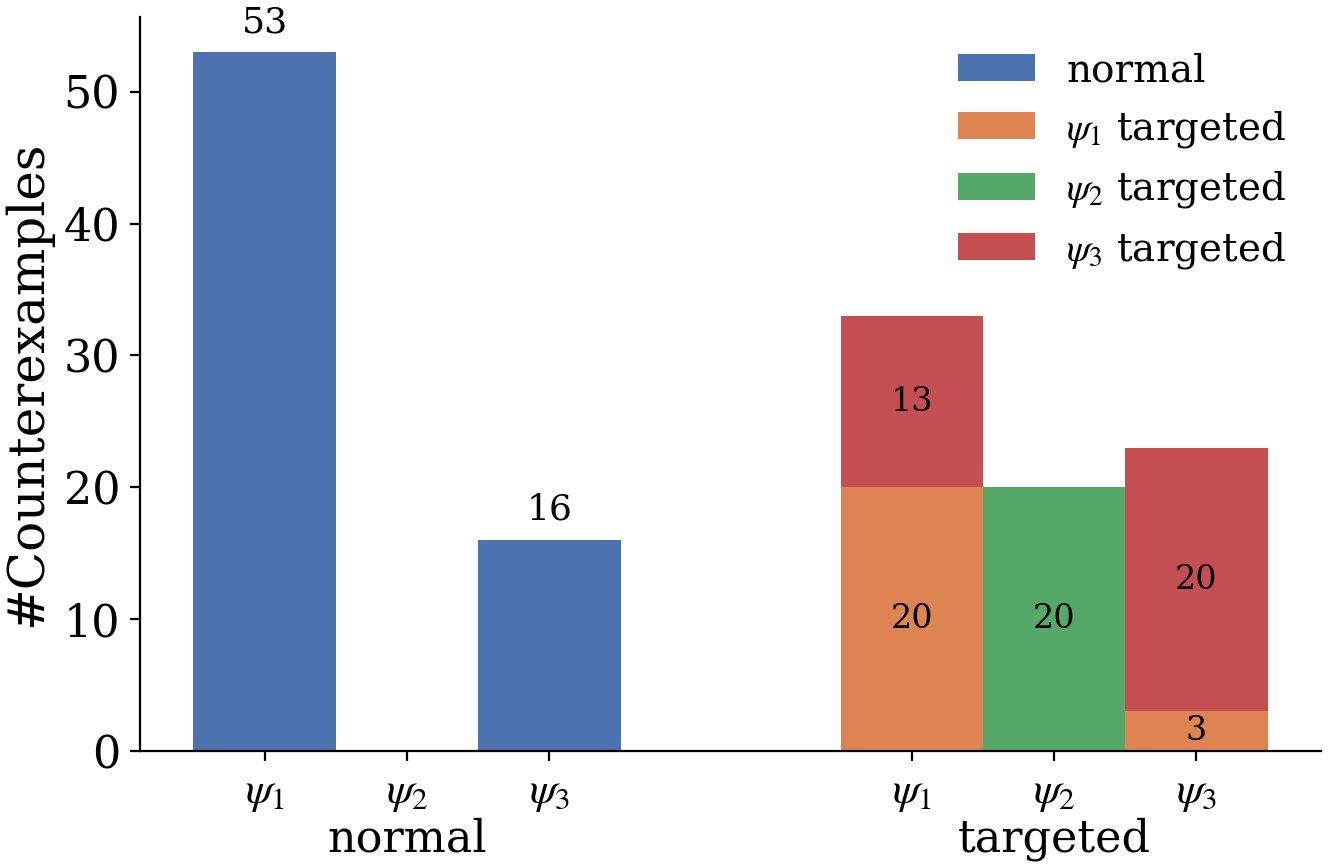}
    \caption{Comparison between normal falsification and falsification guided by generated classes}
    \label{fig:falsification}
\end{figure}
Our results are presented in Fig.~\ref{fig:falsification}. Note that since each counterexample can belong to multiple classes, the total number of counterexamples found by each approach may exceed the number of falsification trials. By the results, we can observe that although normal falsification is able to falsify $\spec{AT}{\fals}$, mostly it tends to return counterexamples that violate $\psi_1$, because that is easier (compared to $\psi_2$ and $\psi_3$) to falsify. In particular, it fails to find any counterexample that violates $\psi_2$, regardless of the great number of trials. 

In comparison, the targeted falsification guided by our generated classes, can find more diverse counterexamples that cover all of the three classes. This is useful in practice when engineers need to debug and repair the system, because different class of counterexamples can expose different causes at the component level, and all of them are needed to gain comprehensive understanding about system behavior. By finding diverse counterexamples, we can offer engineers such references to take further actions.

\begin{figure*}[!t]
\centering
\begin{subfigure}{0.23\textwidth}
\centering
\includegraphics[width=\linewidth]{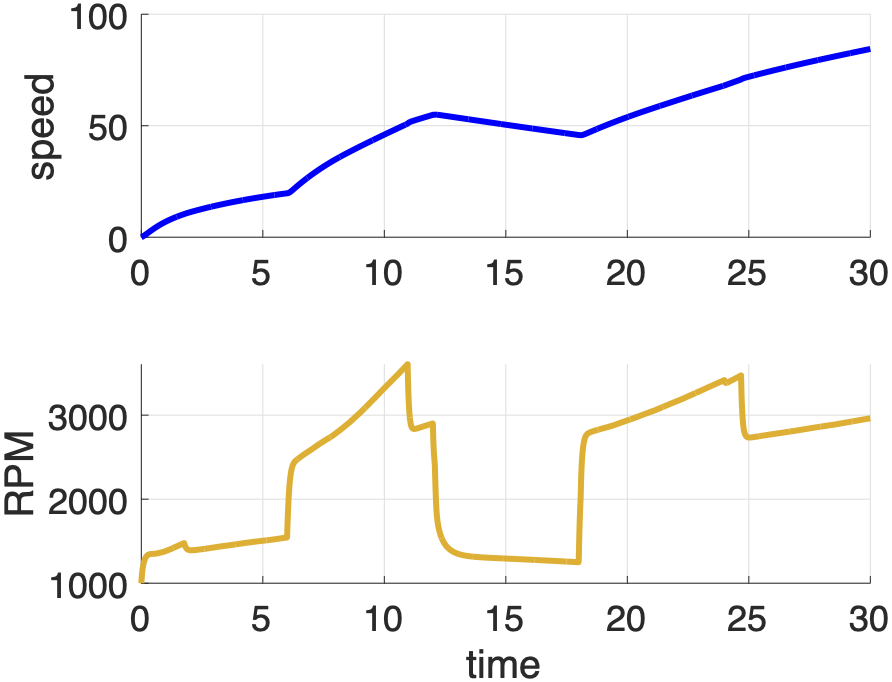}
\caption{The signal}
\label{fig:signal}
\end{subfigure}
\quad
\begin{subfigure}{0.31\textwidth}
\centering
\includegraphics[width=\linewidth]{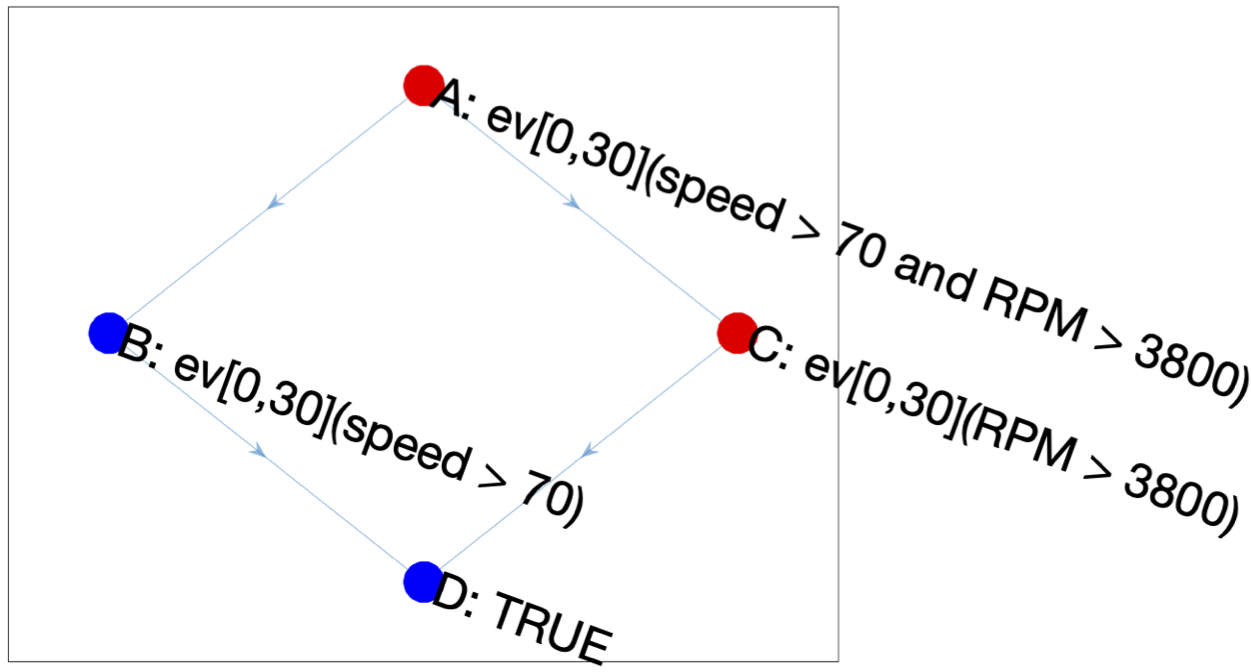}
\caption{Results with $k = 1$}
\label{fig:k1}
\end{subfigure}
\begin{subfigure}{0.42\textwidth}
\centering
\includegraphics[width=\linewidth]{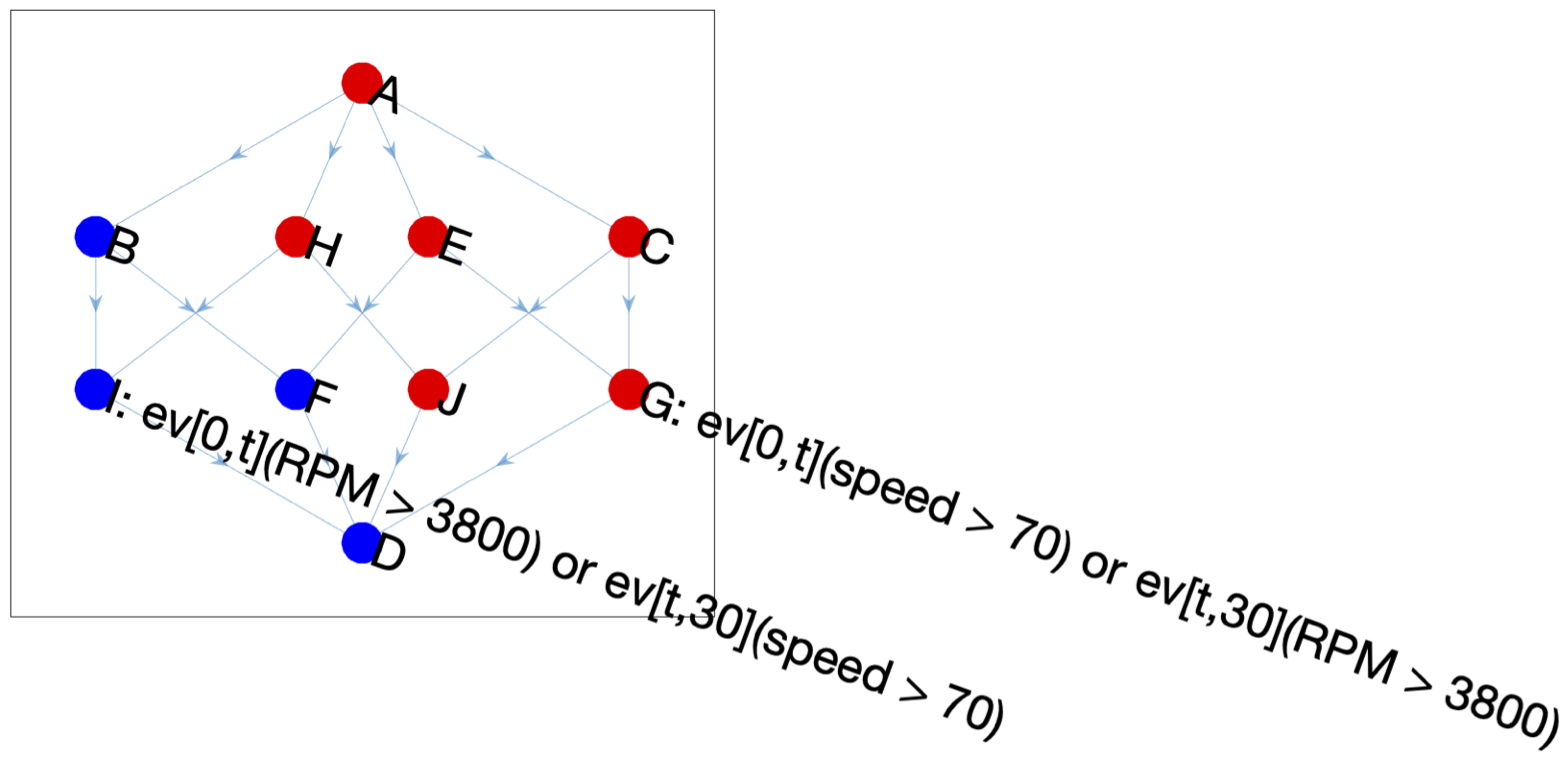}
\caption{Results with $k = 2$}
\label{fig:k2}
\end{subfigure}
\caption{Visualization of classification results for a signal against \spec{AT}{3}. Each node denotes a class, and the edges denote the inclusion relations. Red nodes involve the classes including the signal, while blue nodes are not.}
\label{fig:ourVisual}
\end{figure*}
\myparagraph{Multi-granularity analysis and visualization}
In our criterion, $k$ is a hyperparameter that can control the granularity of classification; by setting different $k$, we can differentiate counterexamples in a finer manner. To demonstrate the effect, we select a counterexample of \spec{AT}{3}, as depicted in Fig.~\ref{fig:signal}, and visualize the results under different $k$s. 

\spec{AT}{3} requires that there must exist a moment when both $\speed>70$ and $\rpm>3800$ happen. 
As shown in Fig.~\ref{fig:signal}, the signal violates the specification in that, while its \speed is greater than 70 at the last few seconds, its \rpm keeps being lower than 3800. This is captured by Fig.~\ref{fig:k1}, because the signal is included in Class C (i.e., $\Diamond_{[0,30]}(\rpm>3800)$) but not in Class B (i.e., $\Diamond_{[0,30]}(\speed>70)$).

Fig.~\ref{fig:k2} visualizes the result when $k=2$, which refines the result of $k=1$ in Fig.~\ref{fig:k1}. For example, we can observe that, the signal is included in 
\begin{math}
    \text{Class G: } \Diamond_{[0,t]}(\speed>70) \lor \Diamond_{[t,30]}(\rpm>3800)
\end{math}, which implies that there exists a pattern that first violates $\speed > 70$ and then violates $\rpm > 3800$; but the signal is not in 
\begin{math}
    \text{Class I: } \Diamond_{[0,t]}(\rpm>3800) \lor \Diamond_{[t,30]}(\speed>70)
\end{math},
which implies that there exists no pattern that violates $\rpm > 3800$ first, and then violates  $\speed > 70$. Notably, this \emph{happen-before} relation between the two events is not encoded in Fig.~\ref{fig:k1} when $k = 1$; therefore, by increasing $k$, we can refine the classification results and deliver more information about the violation of specification.

\section{Related Work}\label{sec:related}

\myparagraph{Monitoring and Falsification} Monitoring~\cite{deshmukh2017robust,bartocci2018localizing,ferrere2015trace,zhang2023online} and falsification~\cite{donze2010breach,AnnpureddyLFS11,ARCHCOMP25Falsification} of hybrid systems have been extensively studied as alternatives to rigorous verification approaches, and especially useful to deal with systems with complex dynamics or   black-box components. However, existing falsification techniques mostly aim to show the existence of counterexamples only, and design their approach to answer the question about whether or not a system can violate given specifications. Existing tools, such as Breach~\cite{donze2010breach}, RTAMT~\cite{nivckovic2020rtamt} and S-TaLiRo~\cite{AnnpureddyLFS11}, can efficiently monitor signals and find counterexamples; advanced tools (such as CPSDebug~\cite{bartocci2021cpsdebug}) can even diagnose the signals to localize the causal part in the signals responsible for the specification violation. However, these frameworks do not support automatic counterexample classification, forcing engineers to manually inspect and analyze the signals to understand their violation patterns. While some approaches~\cite{pedrielli2023part} may be able to detect multiple counterexamples, they do not explicitly classify counterexamples for the purpose of system debugging as we do. 

\myparagraph{Specification mining} The main technical ingredient in our approach involves specification mining, which has been extensively explored in the past decade. We refer readers to a recent survey~\cite{bartocci2022survey} for a comprehensive review. These approaches can be divided to two categories, i.e., template-dependent methods and template-free methods, depending on whether they need a template, e.g., expressed by PSTL. In our case, the specification mining technique we adopt is a template-dependent one, because our classes are represented by PSTL which naturally serve as templates.

\myparagraph{Trace diagnostics} Trace diagnostics~\cite{ferrere2015trace, bartocci2018localizing, onlineResetTCAD2022} is a monitoring technique that can find the causal part of a signal that leads this signal to violation of a given STL specification. It has been applied  in system debugging~\cite{bartocci2021cpsdebug}, and inspires the development of informative monitors~\cite{zhang2023online, zhang2024caumon} that can monitor the causal information at runtime. Our approach is in line with these works, in the sense that, by classification, we also aim to obtain the sufficient condition for STL specifications to be satisfied or violated,  and so these conditions can somewhat be treated as causal information. However, our approach is different from these existing ones, because we do not aim to localize the signal segments that cause the violation; instead, we aim to automatically categorize signals, such that we can clearly present the distribution of counterexamples with respect to different patterns.

\myparagraph{Classification and specification of system behaviors}
There is a growing number of works that aim to classify or specify system behaviors. One line of work~\cite{ferrere2019interface, zhang2019multi, zhang2021effective} is for the purpose of falsification. In~\cite{ferrere2019interface}, Ferr{\`e}re et al. aim to search for counterexamples that are vacuity-free, and so they propose interface-aware STL that can better characterize user requirements. In~\cite{zhang2019multi, zhang2021effective}, Zhang et al. solve the masking problem in falsification by leveraging the syntax tree of STL formulas, however, those approaches cannot differentiate different counterexamples by their violation patterns. There are also works~\cite{bombara2021offline, karagulle2022classification} that aim to classify system behaviors, but these works are essentially specification mining that use STL formulas as classifiers to differentiate different behaviors, rather than classify a set of counterexamples to understand their patterns as we do. Moreover,~\cite{bartocci2020mining} aims to characterize system and summarize their patterns, and specifically they propose \emph{shape expressions} as the formalism for doing that. This approach emphasizes on the shape of signals, rather than different patterns of violation like our approach. Several methods have been developed for classifying or explaining counterexamples generated through model checking of discrete automata or programs~\cite{vick2024counterexample,chechik-counter,beer2012explaining,ball2003symptom}; while the motivation is similar, technically our approach is very different from those works because we consider black-box hybrid systems and our classification criterion relies on specification formalism. Another related work is~\cite{kapoor2024safe}, in which they propose a decomposition method for STL formulas and apply it for safe trajectory planning of autonomous systems. However, due to the different motivation and application domain, their technique cannot be directly used for classification of counterexamples as we do.

 \section{Conclusion and Future Work}\label{sec:conclusion}
 We propose a classification criterion and a framework for the counterexample classification of hybrid systems, enabling us to understand the distribution of counterexamples with respect to different violation patterns. 

As future work, there could be several potential applications of our approach, such as falsification. Existing falsification mostly emphasizes showing the existence of counterexamples, but ignores their diversity. To better assist subsequent system reengineering, our classification technique can be incorporated in the falsification loop to help in searching for more diverse counterexamples.

\bibliographystyle{IEEEtran}
\bibliography{biblio}

\appendices
\section{Proof for Theorem 1}\label{app:Theo1}

\begin{proof}
Theorem~\ref{lem:sound} can be proved by induction on the structure of STL formulas.  First, we show that it holds when $\varphi$ is an atomic proposition $\alpha$ and for the case of satisfaction. In this case, $\psi$ is either $\bot$ or $\alpha$, and
\begin{inparaenum}[1)]
    \item $\bw\models\bot \implies \bw\models\alpha$ holds since it never holds $\bw\models\bot$;
    \item $\bw\models\alpha \implies \bw\models\alpha$ holds obviously.
\end{inparaenum} 
Similarly, it also holds for the violation case. 
    
    Then, we assume that Theorem~\ref{lem:sound} holds for an arbitrary formula $\varphi$, i.e., 
    \begin{inparaenum}[(i)]
        \item\label{assump:1} $\exists\theta.\bw \models\psi(\theta)$ implies $\bw\models\varphi$, where $\psi\in\SCondPlus(\varphi)$, and 
        \item\label{assump:2} $\exists\theta.\bw \not\models\psi(\theta)$ implies $\bw\not\models\varphi$, where $\psi\in\SCondMinus(\varphi)$;
    \end{inparaenum}
    we prove that Theorem~\ref{lem:sound} holds for $\Phi$, which is constructed by applying the operators of STL to $\varphi$. We showcase the following three operators for satisfaction classes and other operators can be derived in a similar manner:
    \begin{compactitem}
            \item $\Phi = \neg\varphi$: In this case, our goal is to show that if it holds $\exists \theta.\bw \models \psi(\theta)$ (where $\psi \in \SCondPlus(\neg\varphi)$), then it holds $\bw \models \neg\varphi$. 
            By STL semantics, given the precondition $\exists\theta.\bw\models\psi(\theta)$ (where $\psi\in \SCondPlus(\neg\varphi)$), it holds that  $\exists\theta.\bw\not\models\neg\psi(\theta)$. By Def.~\ref{def:classCriterion}, since $\psi\in \SCondPlus(\neg\varphi)$, we have $\neg\psi \in \SCondMinus(\varphi)$.  Now we use another symbol $\sigma$ to replace $\neg\psi$, so we can rewrite these conditions as  
            $\exists\theta.\bw \not\models\sigma(\theta)$ (where $\sigma\in \SCondPlus(\varphi)$); by Assumption~(\ref{assump:2}) for $\varphi$, this implies that $\bw\not\models\varphi$, i.e., $\bw\models\neg\varphi$ which is our proof goal in this case, so Theorem~\ref{lem:sound} holds for this case.
        \item $\Phi = \varphi_1\land\varphi_2$: In this case, our goal is to show that if it holds $\exists \theta. \bw \models \psi(\theta)$ (where $\psi\in \SCondPlus(\varphi_1\land \varphi_2$)), then it holds $\bw \models\varphi_1\land\varphi_2$. By Def.~\ref{def:classCriterion}, given $\exists\theta.\bw\models\psi(\theta)$ where $\psi\in\SCondPlus(\varphi_1\land\varphi_2)$, we have $\exists\theta. \bw\models\psi_1(\theta)\land \psi_2(\theta)$, which implies that it holds both $\exists\theta.\bw \models \psi_1(\theta)$ and $\exists\theta.\bw\models\psi_2(\theta)$. By Assumption~(\ref{assump:1}) for $\varphi_1$ and $\varphi_2$, and by Def.~\ref{def:classCriterion}, $\exists\theta.\bw\models\psi_1(\theta)$ implies that $\bw\models\varphi_1$, and $\exists\theta.\bw\models\psi_2(\theta)$ implies that $\bw\models\varphi_2$, so it holds that $\bw\models \varphi_1\land\varphi_2$, which is our proof goal in this case, so Theorem~\ref{lem:sound} holds for this case.
        
        \item $\Phi = \Box_{[a,b]}\varphi$: In this case, our goal is to show that if it holds $\exists\theta. \bw\models\psi(\theta)$ (where $\psi \in \SCondPlus(\Box_{[a,b]} \varphi)$), then it holds $\bw\models \Box_{[a,b]}(\varphi)$. By Def.~\ref{def:classCriterion}, given $\exists\theta.\bw\models\psi(\theta)$ where $\theta\in\SCondPlus(\Box_{[a,b]}\varphi)$, we have $\exists\theta.\bw\models\bigwedge_{i\in\kset{k}}\Box_{[l_i, u_i]}\psi_i(\theta)$, where $\psi_i\in\SCondPlus(\varphi)$. This can further imply that, for any given $t\in [a, b]$, there exists $\psi\in\SCondPlus(\varphi)$ such that $\exists\theta.\bw\models \psi_i$. By Assumption~(\ref{assump:1}), this implies that for any given $t\in[a,b]$, $\bw\models\varphi$, i.e., $\bw\models\Box_{[a,b]}\varphi$, so Theorem~\ref{lem:sound} holds for this case.
    \end{compactitem}
\end{proof}

\section{Proof for Theorem 2}\label{app:theo2}

\begin{proof}
    Similarly to Theorem~\ref{lem:sound}, the proof is by induction. First, Theorem~\ref{lem:complete} holds for the atomic propositions obviously. 

    Then, we assume that Theorem~\ref{lem:complete} holds for an arbitrary formula $\varphi$, i.e., if $\bw\models\varphi$, then there exists $\psi\in \SCondPlus(\varphi)$ s.t. $\bw\in\traceSat{\psi}$, and if $\bw\not\models\varphi$, then there exists $\psi\in \SCondMinus(\varphi)$ s.t. $\bw\in\traceVio{\psi}$; we prove that it also holds for $\Phi$, which is constructed by applying the operators of STL to $\varphi$. We showcase the following three operators for satisfaction classes and other operators can be derived in a similar manner:

    \begin{compactitem}
        \item $\Phi = \neg\varphi$: Our goal is to prove that if $\bw \models \neg\varphi$, then there exists $\psi\in \SCondPlus(\neg\varphi)$ s.t. $\bw\in \traceSat{\psi}$. Given $\bw\models\neg\varphi$, we have $\bw\not\models\varphi$; by assumption, it implies that there exists $\psi\in \SCondMinus(\varphi)$ s.t. $\bw\in\traceVio{\psi}$. By Def.~\ref{def:classCriterion}, we have $\neg\psi\in \SCondPlus(\neg\varphi)$, and by Def.~\ref{def:volume}, $\bw\in\traceVio{\psi}$ implies $\bw\in \traceSat{\neg\psi}$. Therefore, $\neg\psi$ satisfies that $\neg\psi\in \SCondPlus(\neg\varphi)$ and $\bw \in\traceSat{\neg\psi}$, thereby proving Theorem~\ref{lem:complete} for this case.
        \item $\Phi = \varphi_1\land \varphi_2$: Our goal is to prove that if $\bw \models \varphi_1 \land \varphi_2$, then there exists $\psi\in \SCondPlus(\varphi_1\land\varphi_2)$ s.t. $\bw\in \traceSat{\psi}$. Given $\bw \models \varphi_1$ and $\bw \models \varphi_2$, by assumption, there exist $\psi_1\in \SCondPlus(\varphi_1)$ and $\psi_2\in \SCondPlus(\varphi_2)$ such that $\bw\in \traceSat{\psi_1}$ and $\bw\in \traceSat{\psi_2}$. By Def.~\ref{def:classCriterion}, we have $\psi_1\land \psi_2\in \SCondPlus(\varphi_1\land \varphi_2)$. By Def.~\ref{def:volume}, there exist valuations $\theta_1$ and $\theta_2$ such that $\bw\models\psi_1(\theta_1)$ and $\bw\models \psi_2(\theta_2)$; as $\theta_1$ and $\theta_2$ are valuations for different variables, their composition ensures $\bw\in \traceSat{\psi_1 \land \psi_2}$. Given  $\psi_1\land \psi_2\in \SCondPlus(\varphi_1\land \varphi_2)$ and $\bw\in \traceSat{\psi_1 \land \psi_2}$,  Theorem~\ref{lem:complete} holds in this case.

        \item $\Phi = \Box_{[a,b]}\varphi$: Our goal is to prove that if $\bw \models \Box_{[a,b]}\varphi$, then there exists $\psi\in \SCondPlus(\Box_{[a,b]}\varphi)$ s.t. $\bw\in \traceSat{\psi}$. Given $\bw \models \Box_{[a,b]}\varphi$ and the parameter $k$, we can split the interval $[a,b]$ into $k$ segments and we have $\bw\models\bigwedge_{i\in\kset{k}}\Box_{[l_i, u_i]}\varphi$; by assumption, there exists $\psi_i$ ($i\in\kset{k}$) such that $\bw\models\bigwedge_{i\in\kset{k}}\Box_{[l_i, u_i]}\psi_i(\theta_i)$.  Note that, each $\psi_i$ captures a sufficient condition satisfied by $\bw$ in $[l_i, u_i]$, so there could be different ways of splitting, which result in different $\psi_i$ ($i\in\kset{k}$). In practice, the splitting may pursue making each $\psi_i$ capture a \emph{proper subset} of $\traceSat{\varphi}$, however, as $\varphi\in \SCondPlus(\varphi)$, despite how $[a,b]$ is split, $\varphi$ is always an option of $\psi_i$. By Def.~\ref{def:classCriterion}, we have $\bigwedge_{i\in\kset{k}}\Box_{[l_i, u_i]}\psi_i \in \SCondPlus(\Box_{[a,b]}\varphi)$. By Def.~\ref{def:volume}, similar to the conjunction case, we have $\bw\in \traceSat{\bigwedge_{i\in\kset{k}}\Box_{[l_i, u_i]}\psi_i}$. So, Theorem~\ref{lem:complete} holds in this case. 
    \end{compactitem}
\end{proof}

\section{Proof Sketch for Lemma 1}\label{app:Lem1}

\noindent{\it Proof sketch.}
    The proof can be done by induction. As this lemma is not our main conclusion, we sketch the proof for conjunction only to showcase the idea.

    The induction assumption includes that, given $\psi\in \SCondPlus(\varphi)$ and $\sigma\in \SCondPlus(\varphi)$, it holds $\psi\lor \sigma \in \SCondPlus(\varphi)$. We discuss the case $\Phi = \varphi_1\land \varphi_2$, where both $\varphi_1$ and $\varphi_2$ satisfy the induction assumption, and we show that for any $\psi'\in \SCondPlus(\varphi_1\land \varphi_2)$ and $\sigma' \in \SCondPlus(\varphi_1\land \varphi_2)$, $\psi'\lor \sigma' \in \SCondPlus(\varphi_1\land \varphi_2)$.

    By Def.~\ref{def:classCriterion}, $\psi' = \psi^\dag \land \psi^\ddag$ and $\sigma' = \sigma^\dag \land \sigma^\ddag $, where $\psi^\dag, \sigma^\dag \in \SCondPlus(\varphi_1)$ and $\psi^\ddag, \sigma^\ddag \in \SCondPlus(\varphi_2)$. Then, we show that $(\psi'\lor \sigma') = (\psi^\dag \land \psi^\ddag) \lor  (\sigma^\dag \land \sigma^\ddag) \in \SCondPlus(\varphi_1\land\varphi_2)$. By expanding it, we obtain the following formula:
    \begin{align}\label{eq:closureProof}
        (\psi^\dag \lor \sigma^\dag) \land (\psi^\dag\lor \sigma^\ddag) \land (\psi^\ddag \lor \sigma^\dag) \land (\psi^\ddag \lor \sigma^\ddag)
    \end{align}
    By assumption, we have $\psi^\dag\lor \sigma^\dag\in \SCondPlus(\varphi_1)$, $\psi^\dag \lor \sigma^\ddag \in \SCondPlus(\varphi_1\lor \varphi_2)$, $\psi^\ddag \lor \sigma^\dag \in \SCondPlus(\varphi_1\lor \varphi_2)$, $\psi^\ddag \lor \sigma^\ddag \in \SCondPlus(\varphi_2)$. Then by the conjunctive case definition in Def.~\ref{def:classCriterion}, we have Eq.~\ref{eq:closureProof} is a class in $\SCondPlus(\varphi_1\land \varphi_2)$. 
\qed

\section{Proof for Theorem 3}\label{app:theo3}

\begin{proof}
    The proof is also inductive. First, for the case of atomic proposition $\alpha$, it holds obviously.

    Then, we assume that Theorem~\ref{lem:identification} holds for an arbitrary formula $\varphi$, i.e., $\pair{\psi}{\sigma}\in \smallerSatEq(\varphi)$ implies that $\traceSat{\psi}\subseteq\traceSat{\sigma}$, and $\pair{\psi}{\sigma}\in\smallerVioEq(\varphi)$ implies that $\traceVio{\psi}\subseteq \traceVio{\sigma}$. We prove that Theorem~\ref{lem:identification} holds for $\Phi$, which is constructed by applying the operators of STL to $\varphi$. We show the following three cases for satisfaction classes and other cases can be proved similarly:
    \begin{itemize}
        \item $\Phi = \neg\varphi$: Our goal is to prove that $\pair{\psi}{\sigma}\in \smallerSatEq(\neg\varphi)$ implies that $\traceSat{\psi}\subseteq\traceSat{\sigma}$. By Def.~\ref{def:orderIdentify}, if $\pair{\psi}{\sigma}\in \smallerSatEq(\neg\varphi)$, then $\pair{\neg\psi}{\neg\sigma}\in\smallerVioEq(\varphi)$; by assumption, it implies that $\traceVio{\neg\psi} \subseteq \traceVio{\neg\sigma}$. By Def.~\ref{def:volume}, we can infer that $\traceVio{\neg\psi} = \traceSat{\psi}$ and $\traceVio{\neg\sigma} = \traceSat{\sigma}$, so it holds $\traceSat{\psi} \subseteq \traceSat{\sigma}$. 
        \item $\Phi = \varphi_1\land\varphi_2$: Our goal is to prove that $\pair{\psi}{\sigma}\in \smallerSatEq(\varphi_1\land\varphi_2)$ implies that $\traceSat{\psi}\subseteq\traceSat{\sigma}$. By Def.~\ref{def:orderIdentify}, if $\pair{\psi}{\sigma} \in \smallerSatEq(\varphi_1\land\varphi_2)$, then $\psi$ can be represented as $\psi_1\land\psi_2$, $\sigma$ can be represented as $\sigma_1\land\sigma_2$, such that $\pair{\psi_1}{\sigma_1}\in\smallerSatEq(\varphi_1)$ and $\pair{\psi_2}{\sigma_2}\in\smallerSatEq(\varphi_2)$. By assumption, these conditions imply that $\traceSat{\psi_1}\subseteq\traceSat{\sigma_1}$ and $\traceSat{\psi_2}\subseteq\traceSat{\sigma_2}$. By Def.~\ref{def:volume}, this implies that, if $\exists\theta.\bw\models\psi_1(\theta)$, then $\exists\theta.\bw\models\sigma_1(\theta)$, and similarly if $\exists\theta.\bw\models\psi_2(\theta)$, then $\exists\theta.\bw\models\sigma_2(\theta)$. So, if $\exists\theta.\bw\models\psi_1(\theta)\land\psi_2(\theta)$, then $\exists\theta.\bw\models\sigma_1(\theta)\land\sigma_2(\theta)$, which implies that $\traceSat{\psi_1\land\psi_2}\subseteq \traceSat{\sigma_1\land\sigma_2}$.
        \item $\Phi = \Box_{[a,b]}\varphi$: Our goal is to prove that $\pair{\psi}{\sigma}\in \smallerSatEq(\Box_{[a,b]}\varphi)$ implies that $\traceSat{\psi}\subseteq\traceSat{\sigma}$. By Def.~\ref{def:orderIdentify}, if $\pair{\psi}{\sigma}\in \smallerSatEq(\Box_{[a,b]}\varphi)$, then $\psi$ can be represented as $\bigwedge_{i\in\kset{k}}\left(\Box_{[l_i, u_i]}\psi_i\right)$, and $\sigma$ can be represented as $\bigwedge_{i\in\kset{k}}\left(\Box_{[l_i, u_i]}\sigma_i\right)$ such that $\pair{\psi_i}{\sigma_i}\in\smallerSatEq(\varphi)$ for any $i\in\kset{k}$. By assumption, it holds that $\traceSat{\psi_i} \subseteq \traceSat{\sigma_i}$ for each $i\in\kset{k}$. By Def.~\ref{def:volume}, this implies that for a given $i$, if $\exists\theta.\bw\models\psi_i(\theta)$, then $\exists\theta.\bw\models\sigma_i(\theta)$. So, if $\exists\theta.\bw\models \Box_{[l_i,u_i]}\psi(\theta)$, then $\exists\theta.\bw\models \Box_{[l_i,u_i]}\sigma(\theta)$; furthermore, if $\exists\theta.\bw\models\bigwedge_{i\in\kset{k}}\Box_{[l_i,u_i]}\psi_i(\theta)$, then $\exists\theta.\bw\models\bigwedge_{i\in\kset{k}}\Box_{[l_i,u_i]}\sigma_i(\theta)$, so it holds $\traceSat{\psi}\subseteq\traceSat{\sigma}$. 
    \end{itemize}
\end{proof}

\section{Proof for Theorem 4}\label{app:theo4}

\begin{proof}
    The proof is inductive. First, if $\varphi$ is an atomic proposition $\alpha$, it is obvious that any of its classes hold the theorem. 
    Then, we assume that Theorem~\ref{theo:relationComplete} holds for an arbitrary formula $\varphi$, i.e., for any classes $\psi, \sigma\in \SCondPlus(\varphi)$, it holds that $\traceSat{\psi}\subseteq \traceSat{\sigma}$ implies $ \pair{\psi}{\sigma} \in\smallerSatEq(\varphi)$ and  for any classes $\psi, \sigma\in \SCondMinus(\varphi)$, it holds that $\traceVio{\psi}\subseteq \traceVio{\sigma}$ implies $ \pair{\psi}{\sigma} \in\smallerVioEq(\varphi)$. We prove that Theorem~\ref{theo:relationComplete} holds for $\Phi$, which is constructed by applying the operators of STL to $\varphi$. We show the following three cases for satisfaction classes and other cases can be proved similarly:
     \begin{itemize}
        \item $\Phi = \neg\varphi$: let $\psi', \sigma' \in \SCondPlus(\Phi) = \SCondPlus(\neg\varphi)$, our goal is to prove that $\traceSat{\psi'}\subseteq\traceSat{\sigma'}$ implies $\psi'\smallerSatEq \sigma'$. First, by Def.~\ref{def:classCriterion}, we have $\neg\psi', \neg\sigma' \in \SCondMinus(\varphi)$; then, given $\traceSat{\psi'}\subseteq\traceSat{\sigma'}$, by Def.~\ref{def:volume}, it holds $\traceVio{\neg\psi'} \subseteq \traceVio{\neg\sigma'}$. Since $\neg\psi', \neg\sigma' \in \SCondMinus(\varphi)$ and $\traceVio{\neg\psi'} \subseteq \traceVio{\neg\sigma'}$, by the violation case of the assumption, it holds that $\neg\psi'\smallerVioEq \neg\sigma'$; then by Def.~\ref{def:orderIdentify}, it holds that $\psi' \smallerSatEq \sigma'$, which achieves the proof goal of this case. Similarly, we can also prove the violation case.
        \item $\Phi = \varphi_1\land\varphi_2$: Our goal is to prove that for any $\psi, \sigma \in \SCondPlus(\varphi_1\land \varphi_2)$, $\traceSat{\psi} \subseteq \traceSat{\sigma}$ implies $\psi \smallerSatEq \sigma$. By Def.~\ref{def:classCriterion}, $\psi$ and $\sigma$ can be rewritten as $\psi = \psi_1\land \psi_2$ and $\sigma = \sigma_1\land\sigma_2$, where $\psi_1, \sigma_1 \in \SCondPlus(\varphi_1)$, and $\psi_2, \sigma_2 \in \SCondPlus(\varphi_2)$. Assuming that $\traceSat{\psi_1\land \psi_2} \subseteq \traceSat{\sigma_1\land \sigma_2}$, there can be the following two cases:
       \begin{itemize}
           \item it holds that $\traceSat{\psi_1} \subseteq \traceSat{\sigma_1}$ and $\traceSat{\psi_2} \subseteq \traceSat{\sigma_2}$: in this case, by the assumptions about $\varphi_1$ and $\varphi_2$, it holds that $\psi_1 \smallerSatEq\sigma_1$ and $\psi_2\smallerSatEq \sigma_2$. By Def.~\ref{def:orderIdentify}, it holds that $\psi_1\land \psi_2 \smallerSatEq \sigma_1\land \sigma_2$, i.e., $\psi \smallerSatEq \sigma$, which proves Theorem~\ref{theo:relationComplete} for this case.
           \item it does not hold $\traceSat{\psi_1} \subseteq \traceSat{\sigma_1}$ (but it holds $\traceSat{\psi_2} \subseteq \traceSat{\sigma_2}$): we construct $\psi_1\lor \sigma_1$; by Lemma~\ref{lem:closure}, it holds $\psi_1\lor\sigma_1\in \SCondPlus(\varphi_1)$. Then by assumption, as it holds $\traceSat{\psi_1} \subseteq \traceSat{\psi_1\lor \sigma_1}$, it holds that $\psi_1\smallerSatEq \psi_1\lor \sigma_1$. By Def.~\ref{def:orderIdentify}, it holds that $\psi_1\land \psi_2 \smallerSatEq (\psi_1\lor \sigma_1) \land\psi_2$.

           Moreover, to ensure $\traceSat{\psi_1\land \psi_2} \subseteq \traceSat{\sigma_1\land \sigma_2}$, all the signals $\bw \in \psi_1 \setminus \sigma_1$ must not be included in $\psi_1\land \psi_2$ (otherwise, $\bw$ is included in $\in \psi_1\land \psi_2$ but not in $\sigma_1\land \sigma_2$, leading to a contradiction), so it holds $(\psi_1\lor\sigma_1)\land \psi_2 = \sigma_1\land \psi_2$, and further, $\psi_1\land\psi_2\smallerSatEq \sigma_1\land\psi_2$. Given $\traceSat{\psi_2} \subseteq\traceSat{\sigma_2}$, we derive $\psi_2\smallerSatEq \sigma_2$, so $\sigma_1\land\psi_2 \smallerSatEq \sigma_1\land \sigma_2$. By Corollary~\ref{lem:partial}, we have $\psi_1\land\psi_2 \smallerSatEq\sigma_1\land\sigma_2$, which proves Theorem~\ref{theo:relationComplete} for this case.
           \item The remaining cases can be proved similarly.
       \end{itemize}
        \item $\Phi = \Box_{[a,b]}\varphi$: Our goal is to prove that for any $\psi, \sigma \in \SCondPlus(\Box_{[a,b]}\varphi)$, $\traceSat{\psi}\subseteq \traceSat{\sigma}$ implies $\psi\smallerSatEq\sigma$, where $\psi$ can be written as $\bigwedge_{i\in\kset{k}} \Box_{[l_i, u_i]}\psi_i$, and $\sigma$ can be written as $\bigwedge_{i\in\kset{k}} \Box_{[l_i, u_i]}\sigma_i$, where $\psi_i \in \SCondPlus(\varphi)$ and $\sigma_i \in \SCondPlus(\varphi)$. Assume $\traceSat{\bigwedge_{i\in\kset{k}} \Box_{[l_i, u_i]}\psi_i} \subseteq \traceSat{\bigwedge_{i\in\kset{k}} \Box_{[l_i, u_i]}\sigma_i}$; for each $i\in \kset{k}$, it must hold that $\traceSat{\Box_{[l_i, u_i]}\psi_i} \subseteq \traceSat{\Box_{[l_i, u_i]}\sigma_i}$,  (otherwise, there exists $\bw$ in ${\bigwedge_{i\in\kset{k}} \Box_{[l_i, u_i]}\psi_i} $ but not in  ${\bigwedge_{i\in\kset{k}} \Box_{[l_i, u_i]}\sigma_i}$) and furthermore, $\traceSat{\psi_i} \subseteq \traceSat{\sigma_i}$. Given this, by the assumption, it holds that $\psi_i\smallerSatEq\sigma_i$ for each $i\in \kset{k}$, and by Def.~\ref{def:orderIdentify}, it holds that $\bigwedge_{i\in\kset{k}} \Box_{[l_i, u_i]}\psi_i \smallerSatEq \bigwedge_{i\in\kset{k}} \Box_{[l_i, u_i]}\sigma_i$, which proves Theorem~\ref{theo:relationComplete} for this case. 
    \end{itemize}
\end{proof}

\end{document}